\documentclass[preprint,10pt]{elsarticle}

\journal{Physica D}
\usepackage{setspace}
\usepackage{amssymb} 
\usepackage{amsmath}
\usepackage{mathrsfs}
\usepackage{stmaryrd}
\usepackage{amsthm}
\usepackage{graphicx}
\usepackage{comment}

\usepackage[colorinlistoftodos]{todonotes}
\usepackage{tikz-cd}   
\usepackage[utf8]{inputenc} % allow utf-8 input
\usepackage[T1]{fontenc}    % use 8-bit T1 fonts
\usepackage{hyperref}       % hyperlinks
\usepackage{url}            % simple URL typesetting
\usepackage{booktabs}       % professional-quality tables
\usepackage{amsfonts}       % blackboard math symbols
\usepackage{nicefrac}       % compact symbols for 1/2, etc.
\usepackage{microtype}      % microtypography
\usepackage{amsfonts}       % blackboard math symbols
\usepackage{nicefrac}       % compact symbols for 1/2, etc.
\usepackage{lingmacros}
\usepackage{tree-dvips} 
\usepackage{nccmath}
\usepackage{mathtools}
\usepackage{bbm}
\usepackage{graphicx}  
\DeclareMathOperator*{\loggrad}{log-grad}

\usepackage{amsmath, amssymb, mathtools, bm, etoolbox}
\usepackage[noend]{algpseudocode}
\usepackage{algorithm}
\usepackage {mathrsfs}
\usepackage{caption}
\usepackage{graphicx}
\usepackage{subcaption}
\usepackage{color}
\usepackage{enumitem}

\newtheorem{thm-defn}[theorem]{Theorem/Definition}

\theoremstyle{definition}

\theoremstyle{remark}

\newcommand{\rvline}{\hspace*{-\arraycolsep}\vline\hspace*{-\arraycolsep}}

\newcommand*{\vertbar}{\rule[-1ex]{0.5pt}{2.5ex}}
\newcommand*{\horzbar}{\rule[.5ex]{2.5ex}{0.5pt}}
\newcommand{\ignore}[1]{}{}
\usepackage{graphicx}
\usepackage{amssymb}

\begin{document}

\begin{frontmatter}


\title{Association between COVID-19 cases and international equity indices}

\author[label1]{Nick James} \ead{nicholas.james@sydney.edu.au}
\author[label2]{Max Menzies}
\address[label1]{School of Mathematics and Statistics, University of Sydney, NSW, Australia}
\address[label2]{Yau Mathematical Sciences Center, Tsinghua University, Beijing, China}

\begin{abstract}
This paper analyzes the impact of COVID-19 on the populations and equity markets of 92 countries. We compare country-by-country equity market dynamics to cumulative COVID-19 case and death counts and new case trajectories. First, we examine the multivariate time series of cumulative cases and deaths, particularly regarding their changing structure over time. We reveal similarities between the case and death time series, and key dates that the structure of the time series changed. Next, we classify new case time series, demonstrate five characteristic classes of trajectories, and quantify discrepancy between them with respect to the behavior of waves of the disease. Finally, we show there is no relationship between countries' equity market performance and their success in managing COVID-19. Each country's equity index has been unresponsive to the domestic or global state of the pandemic. Instead, these indices have been highly uniform, with most movement in March. 

\end{abstract}


\end{frontmatter}

%\linenumbers

%\begin{nolinenumbers}
%------------BODY OF PAPER ------------------
\section{Introduction}
\label{Introduction}
COVID-19 has had an immense social and economic impact on countries around the world, claiming many lives, necessitating business closures, and sending financial markets into disarray. This paper addresses the following question: on a country-by-country basis, what has had the most impact on a country's stock market - its total cumulative cases, the growth in new daily cases, the return of second waves of the disease, or the worldwide state of the pandemic? The goal of this paper is to study the worldwide spread of COVID-19, analyze the various waves of the disease on a country-by-country basis, and show that the financial markets have been unresponsive to all developments in new or cumulative cases after March.

%1) Epidemiology and finance research on COVID.
The pandemic has  prompted a substantial amount of attention and research. Epidemiologists have analyzed the spread of COVID-19 and potential measures of containment \cite{Wang2020, Chinazzi2020, Liu2020, Fang2020, Zhou2020, Dehning2020}, while clinical researchers have explored potential treatments for the disease \cite{Jiang2020, Zu2020, Li2020, Zhang2020, Wang2020_CELL, NEJM2020, Corey2020}. In finance, many studies have observed the impact of COVID-19 on stock markets \cite{Zhang2020finance, He2020, Zaremba2020}, particularly regarding financial contagion \cite{Akhtaruzzaman2020, Okorie2020} and stability \cite{Lahmiri2020}. %DO WE SPLIT HERE 
Within the nonlinear dynamics community, a majority of papers on COVID-19 have used new and traditional techniques to analyze and predict the spread of cases and deaths \cite{Khajanchi2020, Ribeiro2020, Chakraborty2020,James2020_chaos}. Of these, improvements  to SIR models \cite{SIRBallesteros2020,SIRBarlow2020,SIRCadoni2020,SIRComunian2020,SIRNeves2020,SIRVyasarayani2020,SIRWeinstein2020} and power-law models \cite{Beare2020,Manchein2020,Blasius2020,Anastassopoulou2020} have been the most popular. There is an absence of research that studies financial markets in conjunction with the spread of the virus.

For this goal, we use new and existing time series analysis techniques. Existing methods of time series analysis are diverse, including power-law models \cite{Vazquez2006, Gopikrishnan1998, Podobnik2009, Liu1999,Beare2020}, and nonparametric methods such as distance analysis  \cite{Moeckel1997}, distance correlation \cite{Szkely2007,Mendes2018,Mendes2019} and network models \cite{Shang2020, Onnela2004}.  Time series analysis has been widely applied to both finance \cite{Fenn2011, Drod2018, Drod2020, Eisler2006, Valenti2018, Wang2006} and epidemiology \cite{Hethcote2000,Chowell2016}, including COVID-19 \cite{Manchein2020, Machado2020,James2020_chaos}.

We implement two methods of clustering time series, which have been previously used in various financial \cite{Basalto2007, Basalto2008, Mantegna1999} and epidemiological applications, including inflammatory diseases \cite{Madore2007}, airborne diseases \cite{Kretzschmar2009}, Alzheimer's disease  \cite{Alashwal2019},  Ebola \cite{Muradi2015}, SARS \cite{Rizzi2010}, and COVID-19 \cite{Machado2020}. The two methods we use are hierarchical clustering \cite{Ward1963,Szekely2005} and the optimal one-dimensional implementation of K-means, Ckmeans.1d.dp \cite{Wang2011}.

%4) Structure of paper.
In each of the proceeding three sections, we implement time series analysis and clustering for a different goal. In Section \ref{COVID19_case_death}, we use a smoothed dynamic implementation of cluster analysis to track the worldwide spread of COVID-19, particularly the change in structure over time. In Section \ref{International_trajectories}, we apply semi-metrics to sets of turning points to classify countries according to the disease's first, second or third wave behavior. In Section \ref{Equity_Market_Dynamics}, we use a new method to analyze case trajectories and equity markets in conjunction, and show the markets are highly concurrent with each other, not any country's case counts. Section \ref{Conclusion} summarizes our findings regarding the considerably different progression of COVID-19 and equity market trends of 2020.

%--------------------SECTION II ----------------
\section{Cumulative COVID-19 case and death spread}
\label{COVID19_case_death}

In this section, we use a dynamic and smoothed implementation of cluster analysis to study the worldwide spread of COVID-19,  track the relationships between countries' cumulative case and death counts, and detect changes in the structure of the two time series. Our data spans 12/31/2019 to 08/31/2020, a period of $T=245$ days. We restrict attention to countries with more than 10 000 cumulative cases at the end of the data period, leaving $n=92$ countries. We order these countries by alphabetical order and let $x_i(t), y_i(t) \in \mathbb{R}$ be the multivariate time series of cumulative daily cases and deaths, respectively, for $i=1,..., n$ and $t=1,...,T$.

\subsection{Cluster-based methodology for multivariate time series}
\label{sec:clustermethodology}

Following \cite{James2020_chaos}, this analysis proceeds in several steps, which are further explained in \ref{appendix:clusterbased}. First, given the multivariate time series of cases or deaths, we generate a logarithmic distance matrix $D^{(t)}$ between counts $x_i(t)$ at time $t$. That is,  $D^{(t)}$ is an $n \times n$ matrix with entries $D^{(t)}_{ij}=|\log(x_i(t)) - \log(x_j(t))|$. Next, we estimate an appropriate number of clusters to partition the counts $x_1(t),..., x_n(t)$ at each time $t$. We average over several methods from the statistical learning literature \cite{Radchenko2017} to produce an estimator $k_{av}(t)$, and then apply exponential smoothing to produce a smoothed integer value $\hat{k}(t)$. Third, we use the distance matrix $D^{(t)}$ to partition the counts into $\hat{k}(t)$ clusters at each $t$. As our data is one-dimensional, we apply the optimal implementation of K-means specific to one-dimensional data, Ckmeans.1d.dp \cite{Wang2011}. 

\begin{figure}
    \centering
    \begin{subfigure}[h]{0.49\textwidth}
        \includegraphics[width=\textwidth]{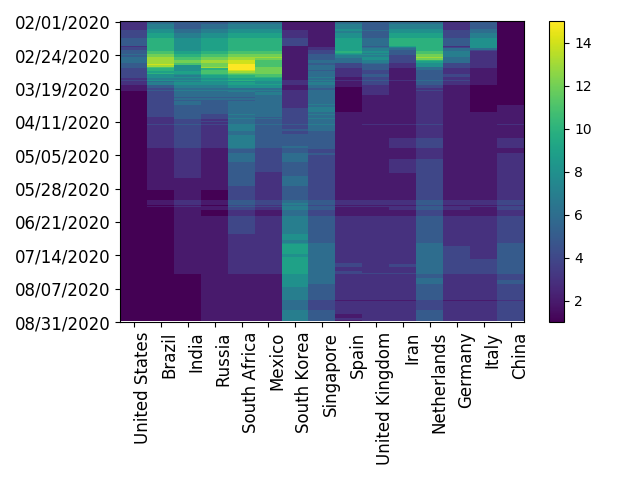}
        \caption{}
        \label{fig:Cluster Membership cases}
    \end{subfigure}
    \begin{subfigure}[h]{0.49\textwidth}
        \includegraphics[width=\textwidth]{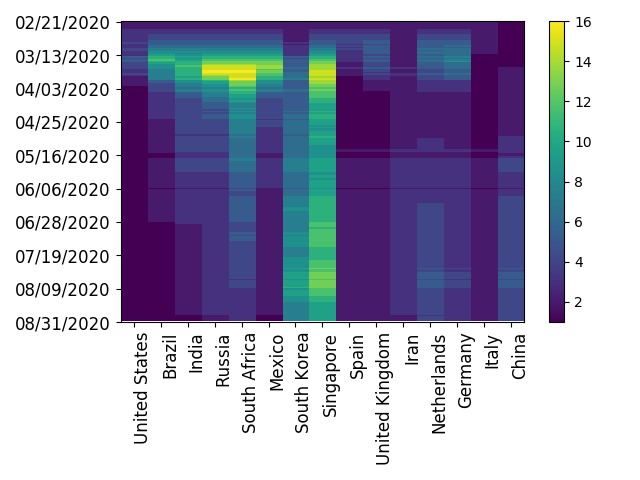}
        \caption{}
        \label{fig:Cluster Membership Deaths}
    \end{subfigure}
    \caption{Heat maps track the changing cluster membership of select countries with respect to (a) cases and (b) deaths, respectively. Cluster membership depicts COVID-19 severity relative to the rest of the world. Clusters are labelled and ordered with 1 being the worst impacted at any time. Darker colors signify worse affected clusters.}
    \label{fig:ClusterMembershps}
\end{figure}

We record the results of this day-by-day clustering in several ways. Figure \ref{fig:ClusterMembershps} displays the changing cluster memberships in the form of heat maps. Figure \ref{fig:Optimal_offset} plots the smoothed number of clusters $\hat{k}(t)$ for both cases and deaths. We define two sequences of $n \times n$ \emph{adjacency matrices} and \emph{affinity matrices} defined by
\begin{align}
\label{eq:adjacency}
    \text{Adj}^{(t)}_{ij} &=
    \begin{cases}
      1 &  x_i(t) \text{ and } x_j(t) \text{ are in the same cluster,}\\
      0, & \text{ else;}
    \end{cases} \\
    \text{Aff}^{(t)}_{ij}&= 1 - \frac{D^{(t)}_{ij}}{\max D^{(t)}}.
    \label{eq:affinity}
\end{align}
To understand the changing cluster structure of the series with time, we define a distance between these adjacency matrices. Let the $L^1$ norm of an $n \times n$ matrix $A$ be defined as $||A||=\sum_{i,j=1}^n |a_{ij}|$. Given $s,t \in [1,..., T]$, let $d(s,t)=||\text{Adj}^{(t)} - \text{Adj}^{(s)}||$. This distance measures the discrepancy between the respective cluster structure on different days. We perform hierarchical clustering on $d(s,t)$ in Figure \ref{datesdendrogram}. %Every construction to this point can be applied to the multivariate time series of either cases or deaths individually.

Finally, we can use the constructions so far to compare the case and death time series in conjunction. Turning first to the number of clusters, let $\hat{k}_X(t), \hat{k}_Y(t)$ be the smoothed number for cases and deaths, respectively. Noticing a similarity between these functions, we can compute the most appropriate offset between them. Given a function $f(t)$ and $\delta>0$, we write $f_\delta (t) = f(t+\delta)$. An appropriate offset can be computed by minimizing the $L^1$ norm between functions, 
\begin{align}
    ||(\hat{k}_X)_{\delta} - \hat{k}_Y||_{L^1} = \int |\hat{k}_X(t+\delta) - \hat{k}_Y(t)| dt.
\end{align}
Turning to the cluster structure, we can define and compute the offset that minimizes the discrepancy between adjacency or affinity matrices $\text{Aff}_X$ and $\text{Aff}_Y$ of the two time series. With $\tau>0$, we seek to minimize the normalized difference
\begin{align}
    \frac{1}{T- \tau} \sum_{t=1}^{T - \tau} ||\text{Aff}^{(t)}_X - \text{Aff}^{(t+\tau)}_Y||.
\end{align}
We display this normalized difference as a function of $\tau$ in Figure \ref{fig:Cluster number evolution} and observe a clear minimum. This can be computed for both adjacency and affinity matrices.

\subsection{Results of cluster-based analysis of cases and deaths}
We now summarize the results of the three figures.  Figure \ref{fig:Cluster Membership cases} tracks the changing cluster membership of 15 select countries with respect to their case counts from February onwards, and captures the natural history of COVID-19.  China was the first country to experience a severe number of cases, and was the unique country in the worst-affected cluster until late March. Then, Italy, Spain and the United States (US) join the worst-affected cluster, struggling to contain their case counts. From the beginning of April until the end of May, the US was the unique member of the worst-affected cluster, signifying how exceptionally it was impacted by COVID-19 cases relative to every other country. Brazil joins the worst-affected cluster at the start of June, and India at the start of August. By contrast, the United Kingdom (UK), Italy, Spain and Germany move to less affected clusters from the beginning of April, likely a result of strict lockdown procedures implemented in these countries.

Figure \ref{fig:Cluster Membership Deaths} tracks the cluster memberships according to deaths for the same countries. Until mid-March, China was the only member of the worst-affected cluster. From then until mid-May, the US, UK, Spain, and Italy belong to the worst-affected cluster. Subsequently, the UK, Spain and Italy leave this cluster. As with cases, the US was the unique member of the worst cluster with respect to deaths for over a month, with Brazil joining at the end of June, and India joining just before the end of August. Given the similarity in case and death cluster behaviors, anomalous countries can be identified if they belong to a significantly different case and death cluster at the same time. The most anomalous country is Singapore. On 08/31/2020, Singapore belonged to the fifth case cluster, but the ninth and least severe death cluster. Indeed, on this day, Singapore had 56771 cases and only 27 deaths, a lower death rate than any other country under consideration. In general, these heat maps have two advantages over other methodologies: first, they are an effective tool for visually comparing the severity of one country's counts relative to the rest of the world. Next, they can highlight countries that are in significantly different clusters with respect to cases and deaths, such as Singapore.

\begin{figure}
    \centering
    \begin{subfigure}[h]{0.49\textwidth}
        \includegraphics[width=\textwidth]{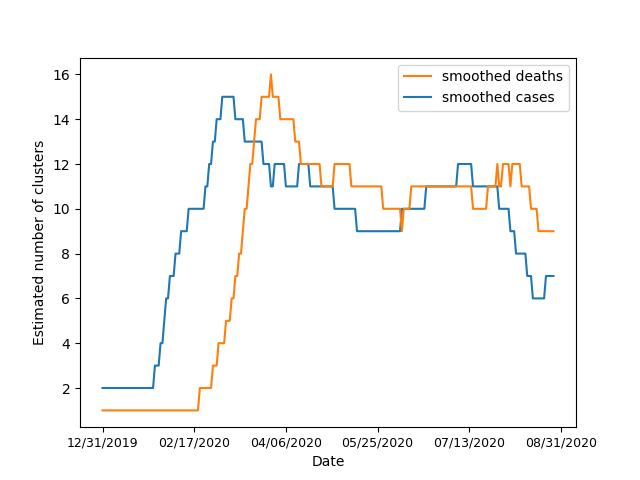}
        \caption{}
        \label{fig:Optimal_offset}
    \end{subfigure}
        \begin{subfigure}[h]{0.49\textwidth}
        \includegraphics[width=\textwidth]{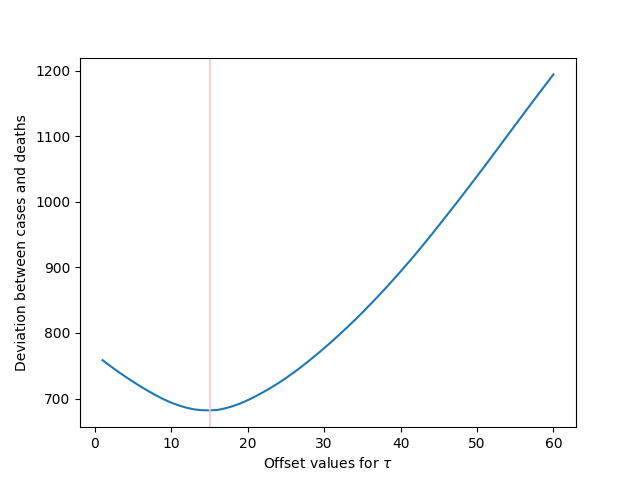}
        \caption{}
        \label{fig:Cluster number evolution}
    \end{subfigure}
    \caption{(a) Smoothed number of clusters $\hat{k}(t)$ as a function of time for both cases and deaths. Similarity is observed up to an offset computed as $\delta=27$. (b) Normalized difference between affinity matrices, with an optimal offset of $\tau=15$.}
\end{figure}

\begin{figure}
\centering
\begin{subfigure}[b]{\textwidth}
\includegraphics[width=\textwidth]{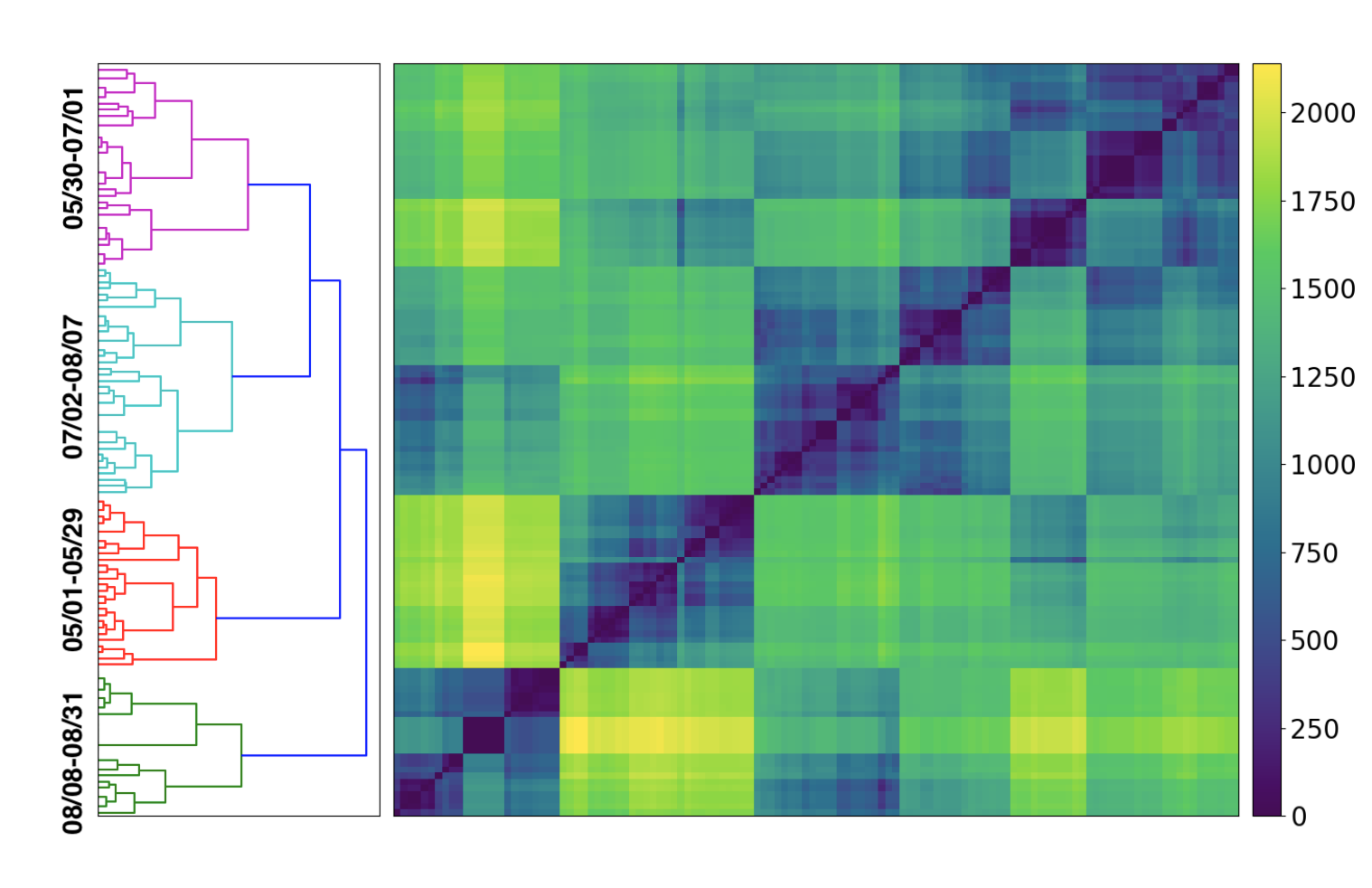}
\caption{}
\label{datesdendrogramcases}
\end{subfigure}
\begin{subfigure}[b]{\textwidth}
\centering
\includegraphics[width=\textwidth]{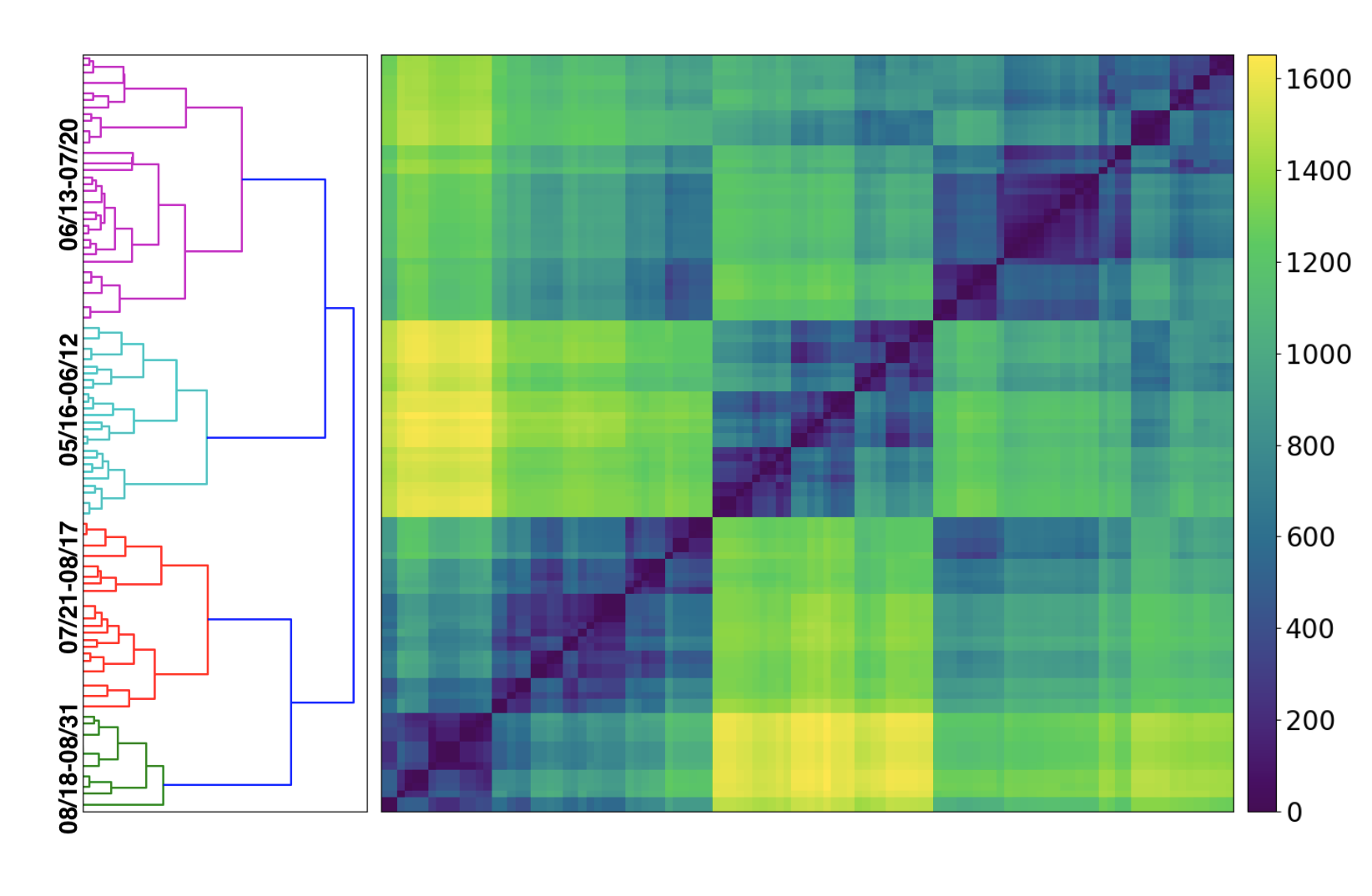}
\caption{}
\label{datesdendrogramdeaths}
\end{subfigure}
\caption{Hierarchical clustering on the distance $d(s,t)$ between adjacency matrices $\text{Adj}^{(t)}$ at different times $t$, for (a) cases and (b) deaths. Each cluster is an unbroken interval of dates. The three boundary dates 05/30, 07/02, 08/08 for cases and 06/13, 07/21, 08/18 for deaths herald significant changes in the structure of the multivariate time series on these dates.}
\label{datesdendrogram}
\end{figure}

Figure \ref{fig:Optimal_offset} tracks the changing number of clusters for both cases and deaths. During February and March, the number of clusters rises substantially as the virus spreads to different countries at different rates. Subsequently, cluster numbers stabilize in April, May and June, then begin to decline as cumulative counts around the world begin to exhibit more homogeneity. Toward the end of the period, the greater number of case clusters than death clusters reflects the greater heterogeneity in death rates than cumulative cases. Singapore is the starkest example here, but this difference reveals a general trend that the time series for deaths become more spread out than the time series for cases. The minimal offset in the number of clusters is computed to be $\delta=27$. Figure \ref{fig:Cluster number evolution} displays a convex minimum of $\tau=15$ for the offset that minimizes the discrepancy between affinity matrices pertaining to the case and death time series. With respect to adjacency matrices, this offset is $\tau=20$.

Figure \ref{datesdendrogram} studies the evolution of the cluster structure over time via hierarchical clustering on the distance between adjacency matrices on different dates. Accounting for the 15-day offset identified in Figure \ref{fig:Cluster number evolution}, we focus our attention on date ranges 05/01 - 08/31 in Figure \ref{datesdendrogramcases} for cases and 05/16 - 08/31 in Figure \ref{datesdendrogramdeaths} for deaths. Each dendrogram identifies four clusters of date ranges, with broad similarity between the figures. All clusters are identified as contiguous intervals of dates: 05/01-05/29, 05/30-07/01, 07/02-08/07, and 08/08-08/31 for cases; 05/16-06/12, 06/13-07/20, 07/21-08/17, and 08/18-08/31 for deaths. Thus, Figure \ref{datesdendrogramcases} reveals marked transitions in the cluster structure on three dates: 05/30, 07/02, and 08/08, while Figure \ref{datesdendrogramdeaths} reveals transitions on 06/13, 07/21, and 08/18. The fact that all clusters are unbroken intervals means the transition dates between the adjacency matrices characterize significant changes in the cluster structure of the respective time series. In Figure  \ref{datesdendrogram}, we plot only the intervals of dates rather than all 123 individual dates, for sake of readability of the labels. This methodology reveals significant changes in a collection of time series without the use of \emph{change point detection algorithms} \cite{Aminikhanghahi2016}. Change point algorithms often require highly tuned parameters for good results, are usually applied in a univariate setting, and typically require the assumption of independence \cite{Lavielle2006}.

%-------------------SECTION III-------

\section{New case trajectories and wave behavior}
\label{International_trajectories}
In this section, we study the trajectories of new cases. Again we restrict to the $n=92$ countries with more than 10 000 total recorded cases as of 08/31/2020. Our goal is to algorithmically identify turning points in new case counts on a country-by-country basis and therefore determine which countries are in their first, second, or later waves of the disease. We also apply a measure of discrepancy between sets of turning points to compare this behavior between countries.

Following \cite{james2020covidusa}, we proceed in several steps, which are further detailed in \ref{appendix:turningpoint}. Let $z_i(t) \in \mathbb{R}$ be the time series of new daily cases, with countries ordered alphabetically. First, we apply a \emph{Savitzky-Golay filter} to produce a collection of smoothed time series $\hat{z}_i(t)$, $t=1,...,T$ and $i=1,...,n$. Next, we apply a two-step algorithm where we select and then refine a set of turning points. We assign to each smoothed time series a non-empty set $P_i$ and $T_i$ of local maxima (peaks) and local minima (troughs). Every sequence of turning points begins with a trough at $t=1$, where there are zero cases, and alternates between trough and peak. These sequences determine if a given country is in a first or second wave of COVID-19. An assigned sequence of TP indicates the country is in its first wave, with case counts that have never materially decreased; a sequence of TPTP indicates a country is in its second wave. A sequence of TPT indicates a country experienced one wave followed by a period of significant decline. If this trough occurs as the last day of the period, the cases are still in decline; if it occurs before the last day, the wave has reached a local minimum and is completely over.

\begin{figure}
    \centering
        \begin{subfigure}[b]{0.32\textwidth}
        \includegraphics[width=\textwidth]{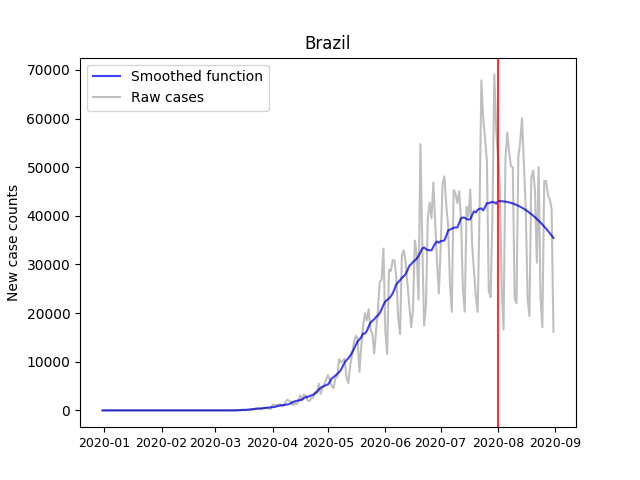}
        \caption{}
        \label{fig:BrazilTS}
            \end{subfigure}
      \begin{subfigure}[b]{0.32\textwidth}
        \includegraphics[width=\textwidth]{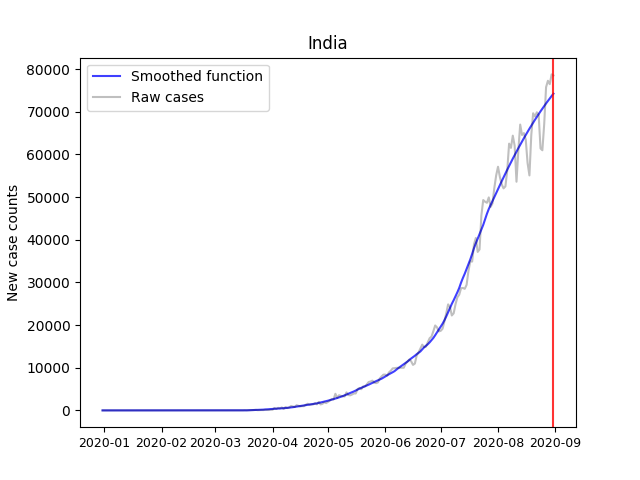}
        \caption{}
        \label{fig:IndiaTS}
    \end{subfigure} 
        \begin{subfigure}[b]{0.32\textwidth}
        \includegraphics[width=\textwidth]{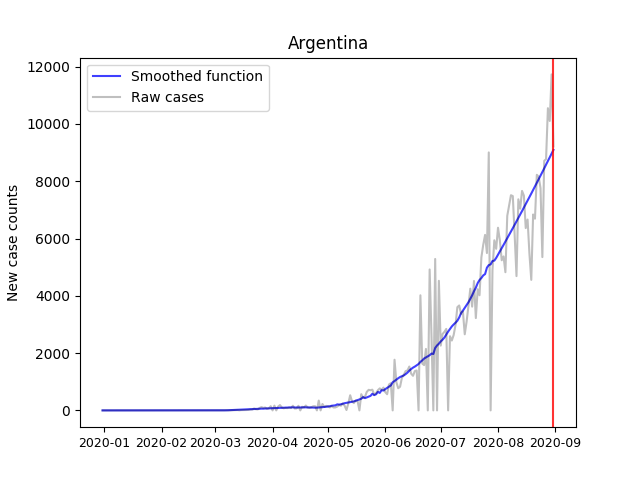}
        \caption{}
        \label{fig:ArgentinaTS}
     \end{subfigure}
     \begin{subfigure}[b]{0.32\textwidth}
        \includegraphics[width=\textwidth]{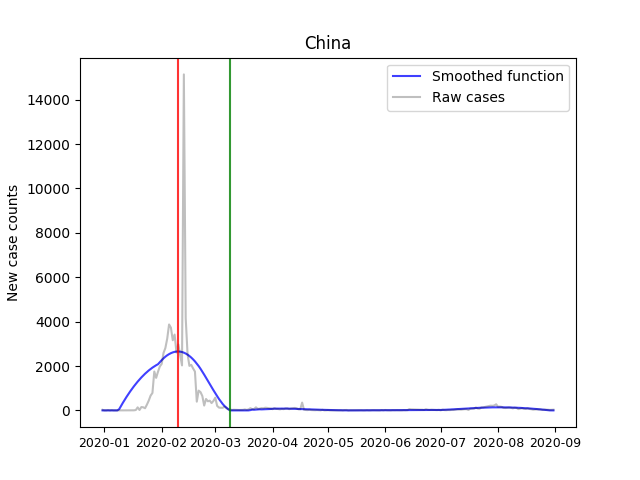}
        \caption{}
        \label{fig:ChinaTS}
     \end{subfigure}
     \begin{subfigure}[b]{0.32\textwidth}
        \includegraphics[width=\textwidth]{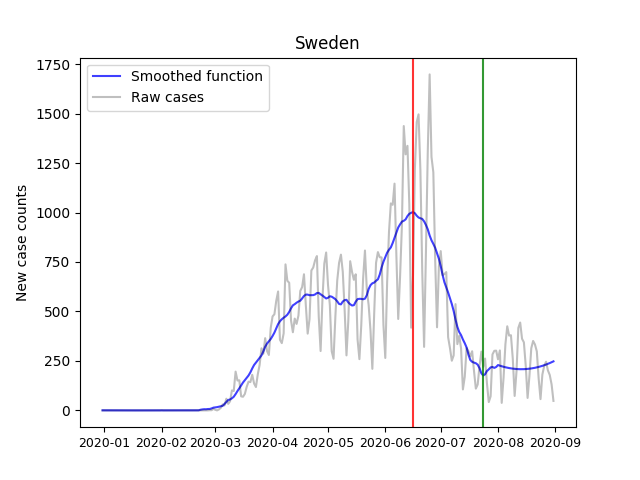}
        \caption{}
        \label{fig:SwedenTS}
    \end{subfigure} 
       \begin{subfigure}[b]{0.32\textwidth}
        \includegraphics[width=\textwidth]{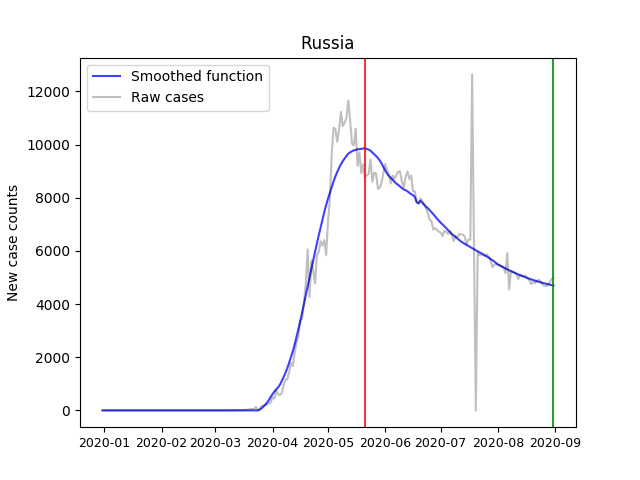}
        \caption{}
        \label{fig:RussiaTS}
    \end{subfigure}
    \begin{subfigure}[b]{0.32\textwidth}
        \includegraphics[width=\textwidth]{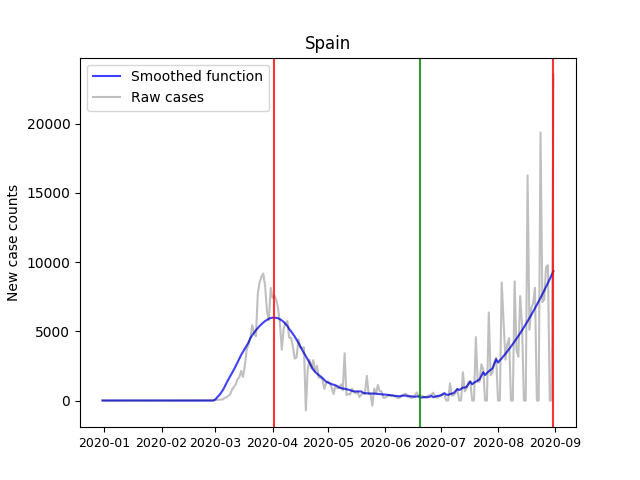}
        \caption{}
        \label{fig:SpainTS}
    \end{subfigure}     
    \begin{subfigure}[b]{0.32\textwidth}
        \includegraphics[width=\textwidth]{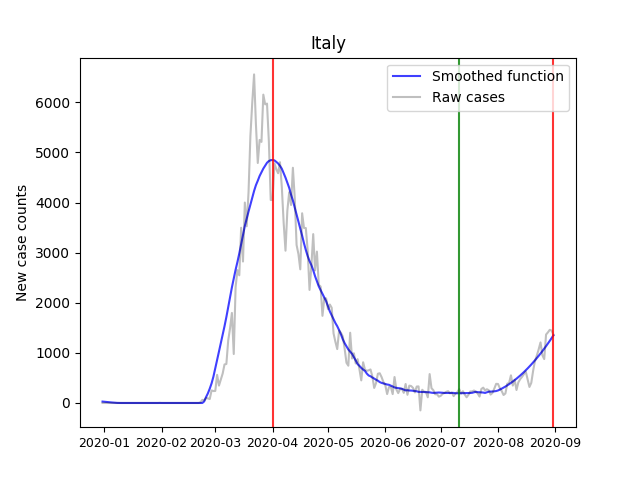}
        \caption{}
        \label{fig:ItalyTS}
    \end{subfigure}    
    \begin{subfigure}[b]{0.32\textwidth}
        \includegraphics[width=\textwidth]{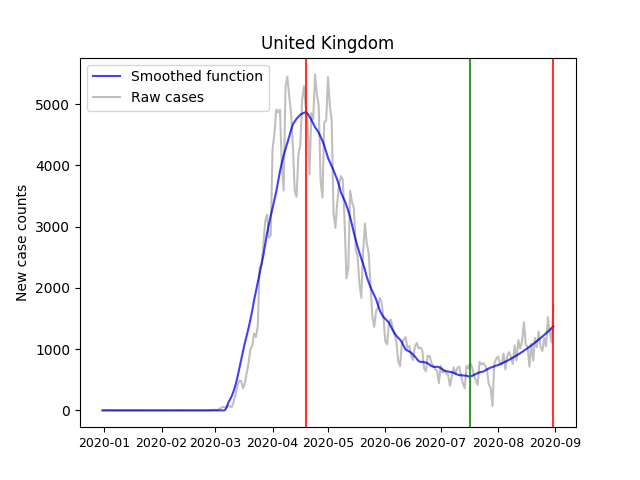}
        \caption{}
        \label{fig:UKTS}
    \end{subfigure}  
    \begin{subfigure}[b]{0.32\textwidth}
        \includegraphics[width=\textwidth]{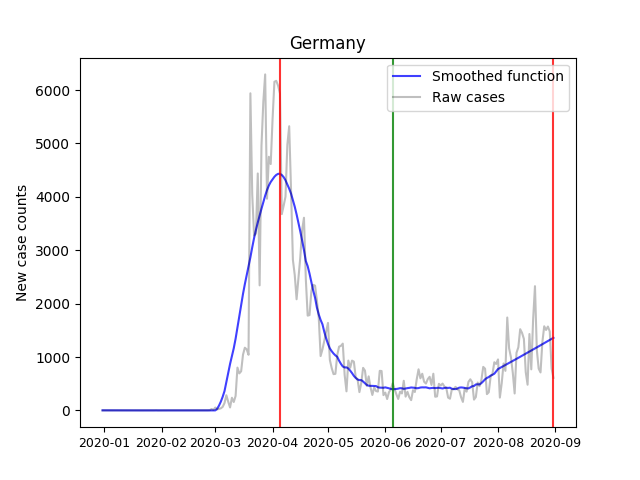}
        \caption{}
        \label{fig:GermanyTS}
    \end{subfigure}     
   \begin{subfigure}[b]{0.32\textwidth}
        \includegraphics[width=\textwidth]{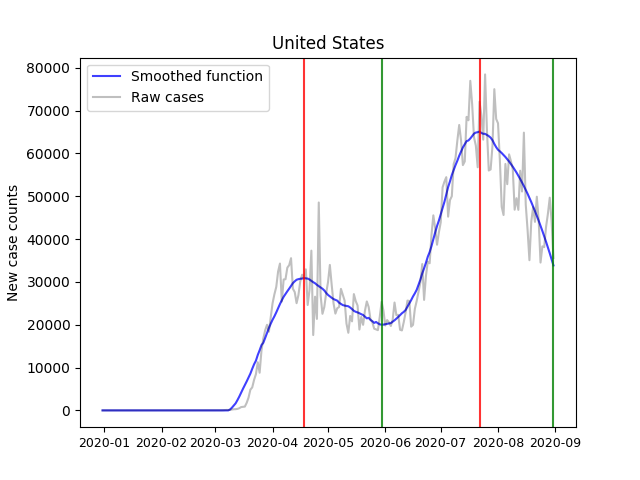}
        \caption{}
        \label{fig:USTS}
    \end{subfigure}     
       \begin{subfigure}[b]{0.32\textwidth}
        \includegraphics[width=\textwidth]{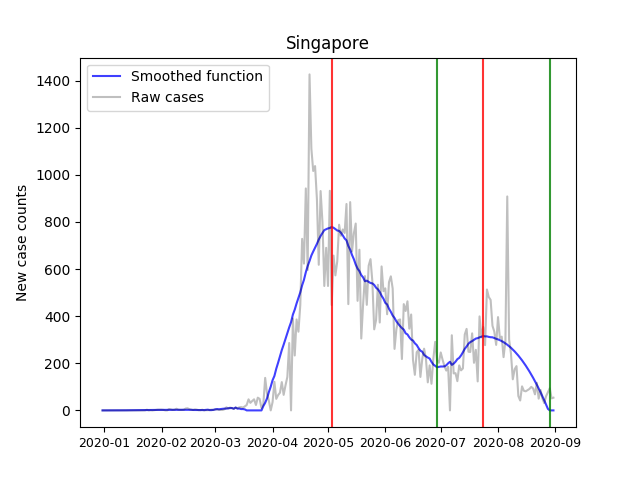}
        \caption{}
        \label{fig:SingaporeTS}
    \end{subfigure}
    \caption{Smoothed time series and identified turning points for various countries: (a) Brazil (b) India  and (c) Argentina are in their first wave. (d) China (e) Sweden and (f) Russia are declining from or have finished their first wave. (g) Spain (h) Italy and (i) the UK are experiencing their second wave. (j) Germany (k) the US and (l) Singapore are declining from their second wave.}
    \label{fig:section2TSplots}
\end{figure}

Finally, we measure distance between two sets of turning points using the semi-metrics between finite sets proposed in \cite{James2020_nsm}. Given two non-empty finite sets $A,B$, this is defined as 
\begin{align}
\label{eq:MJ}
    D({A},{B}) = \frac{1}{2} \left(\frac{\sum_{b\in B} d(b,A)}{|B|} + \frac{\sum_{a \in {A}} d(a,B)}{|A|} \right).
\end{align}
The semi-metric $D(A,B)$ is symmetric, non-negative, and zero if and only if $A=B$. Then, we define the $n \times n$ turning point distance matrix $D^{TP}$ by
\begin{align}
\label{eq:DTPmatrix}
        D_{ij}^{TP} = D(P_i,P_j) + D(T_i,T_j).
\end{align}
Our algorithmic approach classifies the 92 countries under consideration into five characteristic classes. 15 countries, including Brazil, India and Argentina, displayed in Figures \ref{fig:BrazilTS}, \ref{fig:IndiaTS}, \ref{fig:ArgentinaTS}, respectively, are assigned the sequence TP and determined to be in their first wave of the disease. 31 countries, including China (\ref{fig:ChinaTS}), Sweden (\ref{fig:SwedenTS}) and Russia (\ref{fig:RussiaTS}) are assigned the sequence TPT, indicating these countries experienced one wave of the disease - their counts have either reached a local minimum or are still in decline. 28 countries, including Spain (\ref{fig:SpainTS}), Italy (\ref{fig:ItalyTS}), the UK (\ref{fig:UKTS}) and Germany (\ref{fig:GermanyTS}) are assigned TPTP and determined to be in the midst of their second wave. 14 countries, including the United States (\ref{fig:USTS}) and Singapore (\ref{fig:SingaporeTS}) are assigned TPTPT, indicating an ongoing or completed decline from a second wave. Notably, the United States was in a rapid decline in new cases as of 08/31/2020, but still a substantial number of $\sim40 000$, while Singapore's cases declined to nearly zero. Finally, 4 countries, that is, Portugal, Greece, Croatia and South Korea are determined to be in their third wave. Of the 92 countries analyzed, 24 exhibited their greatest case counts up to smoothing on the final day of the period. Of the 46 countries that experienced a second (or third) wave, their final wave was more severe for 28 of them. A complete classification of the 92 countries is included in \ref{sec:classification}.

\begin{figure}
    \centering
    \includegraphics[width=1.2\textwidth]{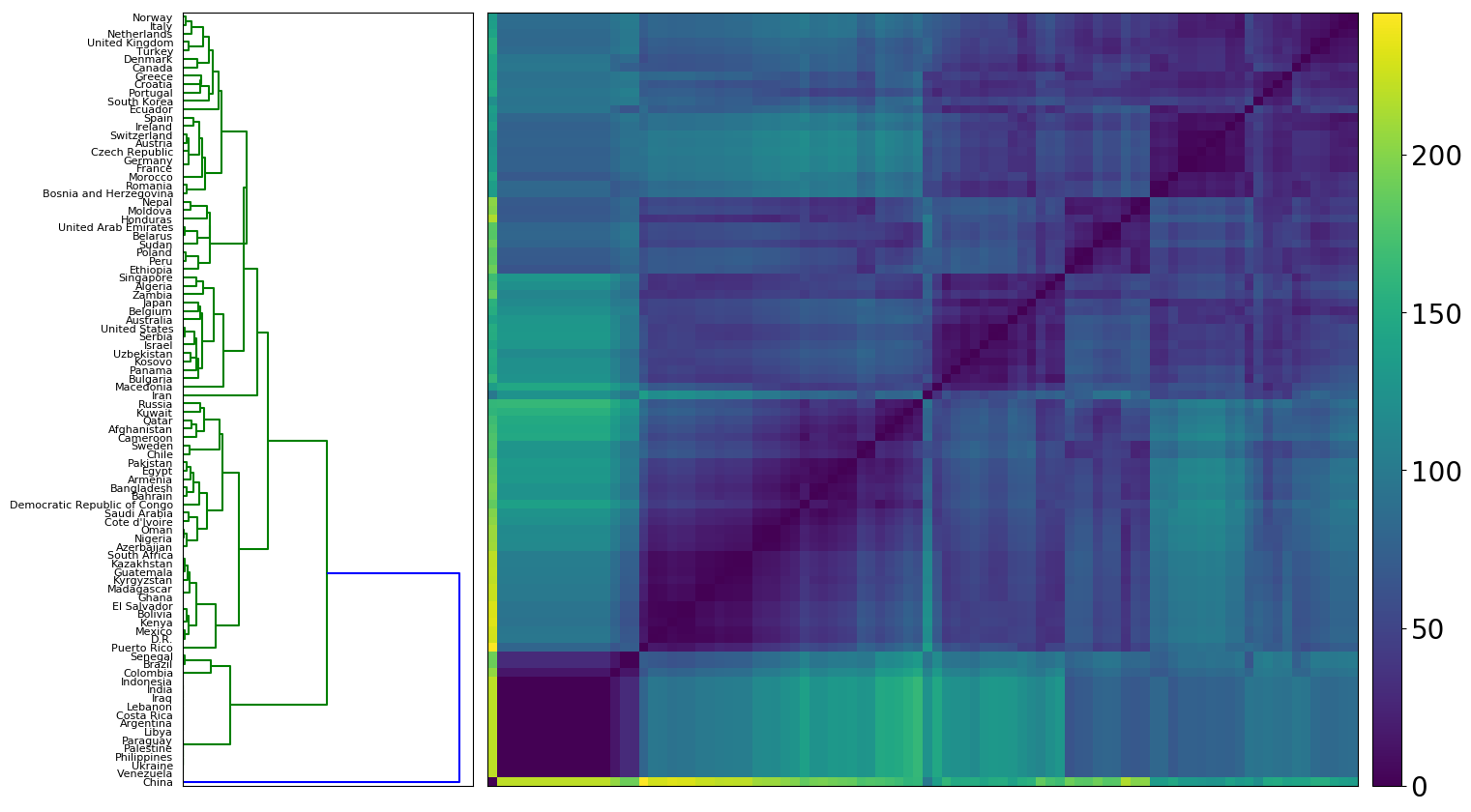}
    \caption{Turning point distance matrix $D^{TP}$, defined in Section \ref{International_trajectories}, measures distance between sets of turning points in new case trajectories. Excluding the outlier China, four primary clusters of time series are identified with the following behaviors: 15 countries in their first wave, 31 countries declining from their first wave, 14 countries declining from their second wave, and a final cluster containing 28 countries currently in their second wave and 4 in their third.}
    \label{fig:Dendrogram_critical_point}
\end{figure}

Figure \ref{fig:Dendrogram_critical_point} displays hierarchical clustering on the $92 \times 92$ matrix $D^{TP}$. China is an outlier due to turning points that occurred much earlier than any other country, and relatively little activity in the disease after March. Excluding China, four primary clusters are revealed in this dendrogram, corresponding to differing behaviors of waves of the disease. The semi-metric in Equation (\ref{eq:MJ}) prioritizes low minimal distances between sets, rather than the number of elements. Thus, the four countries in their third wave are assigned to the same cluster as the second wave countries due to low minimal distances between their turning point sets. Compared to other methodologies, the matrix $D^{TP}$ is capable of quickly classifying countries into first, second or higher wave behaviors. This paper is the first time it has been used to examine different countries' COVID-19 counts.

\section{Equity Market Dynamics}
\label{Equity_Market_Dynamics}
In this section, we study the dynamics of 17 countries' equity indices with respect to both pricing and 30-day rolling volatility. The data spans 01/01/2020 to 08/31/2020, a period of $T_1=175$ trading days. The countries analyzed are: Argentina, Australia, Brazil, Canada, China, France, Germany, India, Italy, Japan, Mexico, Russia, South Korea, Spain, Switzerland, the UK and the US. This list contains the top 15 economies in the world by nominal GDP \cite{worldbankgdp}, and at least one country from each of the five characteristic classes of behavior identified in Section \ref{International_trajectories}.

\begin{figure}
    \centering
    \includegraphics[width=\textwidth]{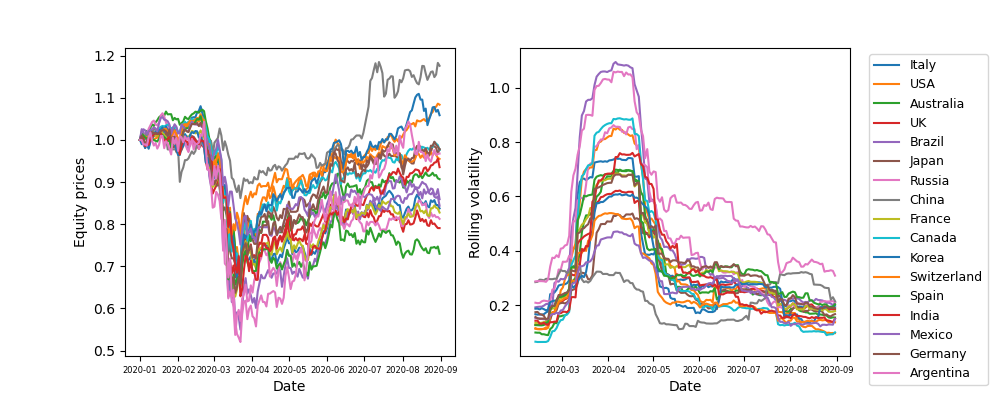}
    \caption{Equity market dynamics for 17 countries with respect to (a) adjusted closing equity prices, normalized to begin the year at 1, and (b) rolling volatility for the prior 30 days.}
    \label{fig:indices}
\end{figure}

Let $p_i(t)$ be the multivariate time series of each country's daily closing equity prices, for $t=1,...,T_1$ and $i=1,..., 17$. Let $\sigma_i(t)$ be the 30-day rolling volatility, $t=1,...,T_1-30$. For each $t$, this is defined as the standard deviation of the previous 30 days of index data, normalized by $\sqrt{250}$, the number of trading days in a year. This measure of volatility is one of several that we could potentially consider. For example, \cite{arouxet2020} study the impact of COVID-19 on cryptocurrencies with a specific analysis of the max-min and absolute return volatility time series. We plot all 17 countries' equity prices and rolling volatility in Figure \ref{fig:indices}, respectively, with equity prices normalized to 1 as of the start of the year. %, and observe significant similarities in their dynamics. 
Every index experiences a significant drop and a highly volatile period in March. At the end of the period, China's index has risen the most relative to its value at the beginning of 2020. Qualitatively, we make two striking observations: first, market dynamics have been highly uniform among the 17 countries, with China as the only exception. Secondly, market movement after March has been largely unaffected by the natural history of COVID-19 described in Sections \ref{COVID19_case_death} and \ref{International_trajectories}, such as Brazil and India entering the worst-affected clusters for cases and deaths by the end of August, or the United States experiencing a large second wave in July, or many other developments.
\begin{comment}
 
\begin{figure}
    \centering
    \begin{subfigure}[b]{0.49\textwidth}
    \includegraphics[width=\textwidth]{Financial_Figures/Share_price_rebased.png}
    \caption{}
    \label{fig:Price_trajectories}
        \end{subfigure}
    \begin{subfigure}[b]{0.49\textwidth}
    \includegraphics[width=\textwidth]{Financial_Figures/Rolling_volatility_computation.png}
    \caption{}
    \label{fig:Volatility_trajectories}
    \end{subfigure}
    \caption{Equity market dynamics for 17 countries with respect to (a) adjusted closing equity prices, normalized to begin the year at 1, and (b) rolling volatility for the prior 30 days.}
    \label{fig:equityindices}
\end{figure}
\end{comment}

We proceed to quantify and further elucidate both of these observations. First, we analyze all trajectories of equity prices and rolling volatilities in conjunction. Considering equity prices of a single country gives a function $\mathbf{p}_{i} \in \mathbb{R}^{T_1}$. Let $||\mathbf{p}_{i}|| = \sum^{T_1}_{t=1} |p_{i}(t)|$ be its $L^1$ norm. We can define a normalized index price trajectory by $\mathbf{g}_{i} = \frac{\mathbf{p}_i}{||\mathbf{p}_i||}$. Analogously, we define $||\mathbf{\sigma}_{i}|| = \sum^{T_1 - 30}_{t=1} |\sigma_{i}(t)|$ and the normalized volatility trajectory by $\mathbf{v}_{i} = \frac{\mathbf{\sigma}_i}{||\mathbf{\sigma}_i||}$. These vectors highlight the relative changes of price or volatility within the entire period. We then define two \emph{trajectory distance matrices}, $D^P_{ij}=||\mathbf{g}_i - \mathbf{g}_j||$ and $D^{\text{vol}}_{ij} = ||\mathbf{v}_i - \mathbf{v}_j||$.

\begin{figure}
    \centering
    \begin{subfigure}[b]{0.45\textwidth}
    \includegraphics[width=\textwidth]{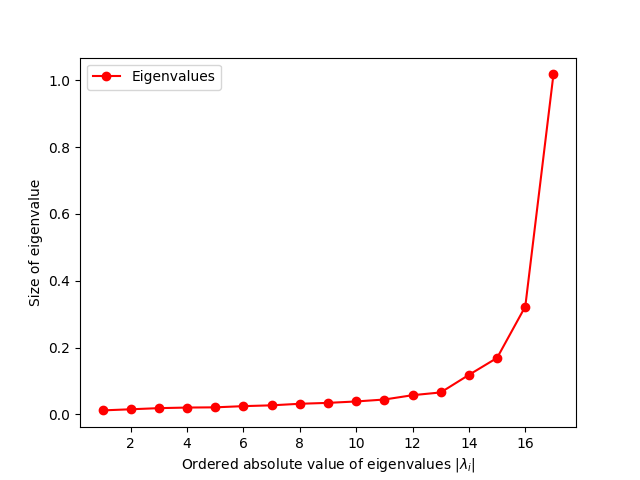}
    \caption{}
    \label{fig:Price_eigen}
        \end{subfigure}
    \begin{subfigure}[b]{0.45\textwidth}
    \includegraphics[width=\textwidth]{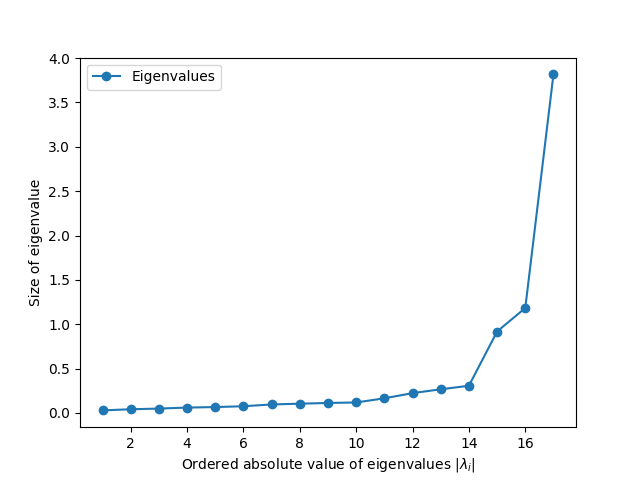}
    \caption{}
    \label{fig:Volatility_eigen}
    \end{subfigure}
    \caption{Absolute value of eigenvalues for the trajectory distance matrices for (a) equity prices and (b) rolling volatility. Choosing $\epsilon=0.2,0.3$ respectively, we detect broad similarity between 16 equity indices with respect to price and 15 with respect to volatility.}
    \label{fig:eigen}
\end{figure}

We analyze these distance matrices $D^P$ and $D^{\text{vol}}$, which are symmetric, real matrices with trace 0. As such, they can be diagonalized with real eigenvalues. To determine self-similarity within these indices with respect to prices and rolling volatility, we plot the absolute values of the eigenvalues $|\lambda_1| \leq ... \leq |\lambda_{17}|$ for each respective matrix in Figure \ref{fig:eigen}. Inspecting the collective similarity of Figure \ref{fig:indices}, we expect that a large number $K$ of the 17 countries are highly similar with respect to equity price and rolling volatility, with a small number of outliers. Indeed, one would expect many volatility trajectories to behave similarly due to structural financial market factors such as volatility clustering. We examine the eigenvalues to estimate this $K$. If there is a large collection of highly similar elements in an $n \times n$ distance matrix $D$, the matrix would have the form
\[
\begin{pmatrix}
  \begin{matrix}
   & c_{1} & c_{2} & c_{3} & \hdots & c_{K} \\
  r_{1} & 0 & * & * & \hdots & * \\
  r_{2} & * & 0 & * & \hdots & * \\
  r_{3} & * & * & 0 & \hdots & *\\
   & \vdots & \vdots & \vdots & \ddots \\
  r_{K} & * & * & * & * & 0 \\
  \end{matrix}
  & \rvline &   
  \begin{array}{ccc}
    \horzbar & r_{1} & \horzbar \\
    \horzbar & r_{2} & \horzbar \\
             & \vdots    &          \\
    \horzbar & r_{K} & \horzbar
  \end{array} & \\
\hline
  \begin{array}{ccccc}
    \vertbar & \vertbar & \vertbar &        & \vertbar \\
    r^{T}_{1}    & r^{T}_{2} & r^{T}_{3}   & \ldots & r^{T}_{K}    \\
    \vertbar & \vertbar & \vertbar &       & \vertbar 
  \end{array} & \rvline &
  \begin{matrix}
  0 &  \\
   & 0 \\
   &  & \ddots \\
   & & & 0
  \end{matrix}
\end{pmatrix}
\]
where rows $r_1,\ldots,r_{K}$ are highly similar to one another and elements $*$ are close to zero. Such a matrix is a small deformation away from a rank $n-(K-1)$ matrix, and so $K-1$ of the eigenvalues should be close to 0.

Thus, given an appropriate threshold $\epsilon$, if $|\lambda_1| \leq ... \leq |\lambda_{K-1}|\leq \epsilon$, then we can deduce $K$ indices are similar with respect to price or volatility. This is a concise measure of the number of indices that are similar within the collection studied. In Figure \ref{fig:Price_eigen}, we can set $\epsilon=0.2$ and observe that 15 eigenvalues are less than this threshold, suggesting high similarity among 16 of the index prices, with China as the clear outlier. In Figure \ref{fig:Volatility_eigen}, 14 eigenvalues are under a threshold of $\epsilon=0.3$, suggesting broad similarity among 15 of the indices with respect to volatility. Returning to Figure \ref{fig:indices}, this broad similarity is even more striking when examining the significant changes over time in market behavior. %which demonstrates the magnitude of change, both increasing and decreasing, in collective behavior across the market.

Further, the eigenvalue analysis provides a measure of the scale of the matrix. Since the distance matrices $D$ are symmetric, they can be conjugated by an orthogonal matrix to yield a diagonal matrix of eigenvalues. As a consequence, the operator norm \cite{RudinFA} of $D$ coincides with the largest eigenvalue $|\lambda_n|$. That is,
\begin{align}
    \max_{x \in \mathbb{R}^n - \{0\}} \frac{||Dx||}{||x||}= ||D||_{\text{op}}= |\lambda_n|.
\end{align}
We can see from Figure \ref{fig:eigen} that the operator norm for $D^{\text{vol}}$ is approximately four times that of $D^P$. Both distance trajectory matrices are normalized, so a direct comparison is appropriate. Similarly, when comparing $L^2$ matrix norms, $||D^{\text{vol}}||_2=4.14$, while $||D^P||_2=1.09$. That is, there is a higher degree of collective similarity among indices with respect to price trajectories than volatility. 
This is a surprising result, given the expected similarity in collective volatility behavior due to volatility clustering. This result may differ if we were to compare price and volatility trajectories of assets in different financial sectors.

Finally, we quantify the extent to which large changes in index price coincide with high case counts. For this purpose, we take all 17 equity price time series $p_i(t)$ and the corresponding countries' new case time series $z_i(t) \in \mathbb{R}_{\geq 0}$. Large values of $z_i(t)$ mean that the disease is spreading rapidly in that country. On the other hand, large values of the absolute value log returns $|R_i(t)|=\left| \log\left( \frac{p_i(t)}{p_i(t-1)} \right)\right|$ indicate significant changes in the value of the market. Since equity data is only applicable on weekdays, we restrict the new case time series to the weekdays to yield a time series $w_i(t), t=1,...,T_1$. As new cases are lower on the weekends, this provides a good representation of the trajectory of new cases in each country.

We define a symmetric $34 \times 34$ matrix $M$ that compares the concurrence of these changes. The entries of $M$ are normalized inner products between time series to measure the extent of overlap between market movement and new cases. We define
\begin{align}
||R_i||_2&=\left(\sum_{t=1}^{T_1} |R_i(t)|^2\right)^\frac{1}{2},\\ ||w_i||_2&=\left(\sum_{t=1}^{T_1} w_i(t)^2\right)^\frac{1}{2},\\
<R_i, w_j>_n&=\frac{1}{||R_i||||w_j||} \sum_{t=1}^{T_1} R_i(t)w_j(t).
\end{align}
The pairing $<.,.>_n$ is a normalized inner product that measures the concurrence of large changes in the time series more accurately than the correlation between price and new case time series. The matrix $M$ is defined as follows:
\begin{align}
M_{ij}=\begin{cases}
<|R_i(t)|,|R_j(t)|>_n \text{ if } 1\leq i,j \leq 17,\\
<w_{i-17}(t),w_{j-17}(t)>_n \text{ if } 18\leq i,j \leq 34,\\
<|R_i(t)|,w_{j-17}(t)>_n \text{ if } 1\leq i \leq 17, 18 \leq j \leq 34.
\end{cases}
\end{align}
As all the sequences $|R_i(t)|,w_j(t)$ are non-negative, all entries of $M$ are non-negative. In general, given non-negative functions $f,g$, $<f,g>_n=1$ if and only if $f=\alpha g$ for some $\alpha>0$, while $<f,g>_n=0$ if and only if the non-zero values of $f$ and $g$ are disjoint sets. 

\begin{comment}
 
Theta notation
\begin{align}
M_{ij}&=\theta(|R_i(t)|,|R_j(t)|) \text{ if } 1\leq i,j \leq 17\\
M_{ij}&=\theta(w_{i-17}(t),w_{j-17}(t)) \text{ if } 18\leq i,j \leq 34\\
M_{ij}&=\theta(|R_i(t)|,w_{j-17}(t)) \text{ if } 1\leq i \leq 17, 18 \leq j \leq 34
\end{align}
Here, $\theta$ is a normalized inner product given by
\begin{align}
    \theta(f(t),g(t))=\frac{\sum_{t} f(t)g(t)}{ (\sum_t f(t)^2)^{0.5}(\sum_t g(t)^2)^{0.5}} 
\end{align}
\end{comment}

\begin{figure}
    \centering
    \includegraphics[width=.85\textwidth]{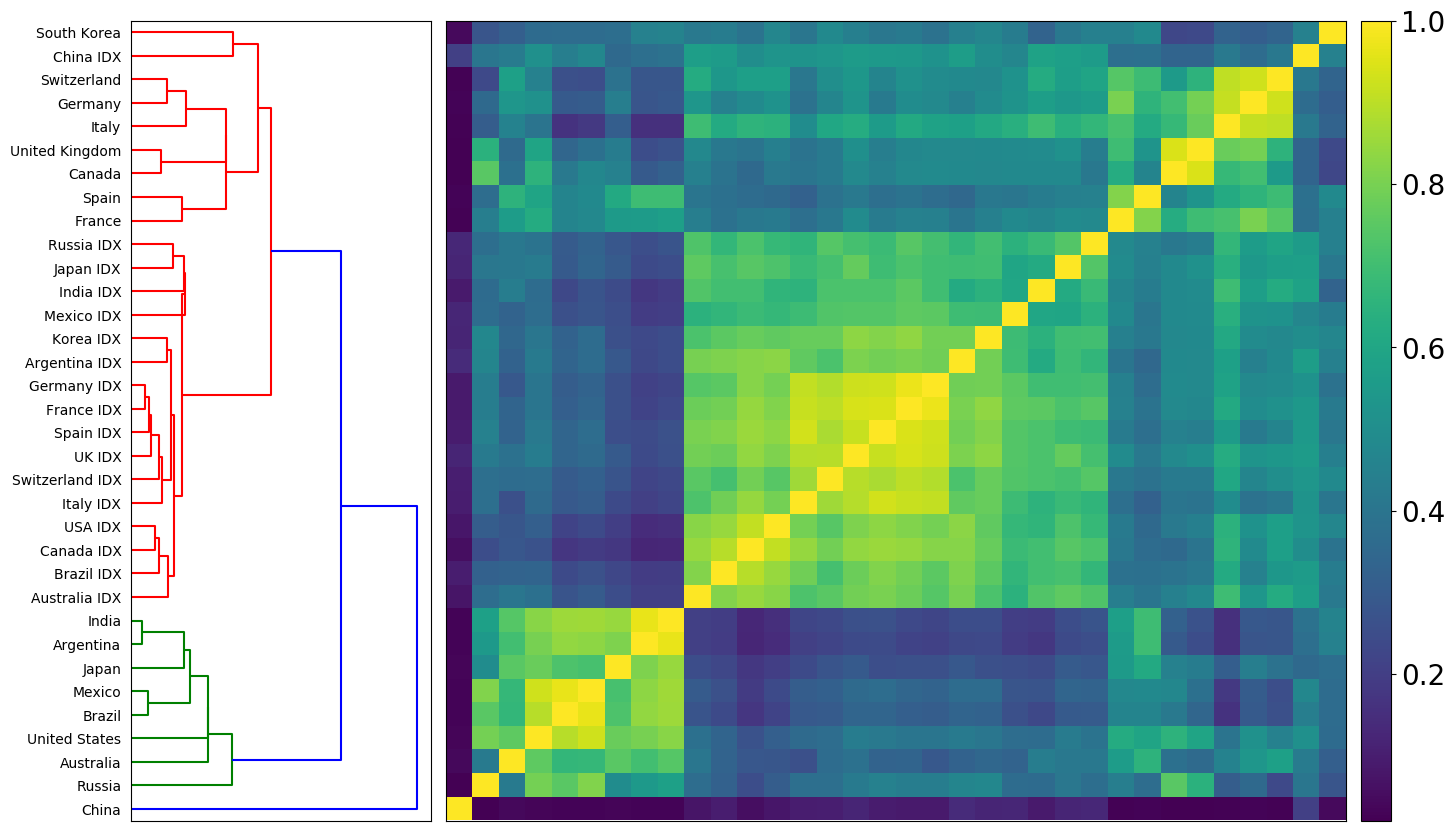}
    \caption{Hierarchical clustering on the normalized inner product matrix $M$. High similarity is observed between 16 equity indices, with no relationship to case counts. China is observed as an outlier in both cases and index. Other countries' case counts split into two subclusters according to whether large counts of new cases occur disproportionately early or later.}
    \label{fig:equitycasesdend}
\end{figure}

In Figure \ref{fig:equitycasesdend} we perform hierarchical clustering on the matrix $M$ and reveal several insights.
First, China is highly anomalous with respect to both case counts and its index. Indeed, China recorded a large number of cases only during January and February, with few cases since and no subsequent wave.
Second, China is also relatively anomalous with respect to its index. We can see two particular periods in Figure \ref{fig:indices} where China did not undergo similar large changes as other countries. In March, China's index experienced a less severe drawdown than every other country; in July, China experienced a period of significant positive growth, unlike any other country.

Third, the dendrogram reveals a high level of similarity among equity indices, excluding China's, visible in the clear subcluster in the center of the dendrogram. These 16 equity indices form a submatrix in which the mean of all the entries is 0.78, indicating high concurrence of large price movements. 
Turning to the remaining indices, the dendrogram reveals more heterogeneity, yet some similarity, between the same 16 countries' case counts. While the 16 equity indices form one prominent subcluster, the same countries' case counts split into two subclusters. The normalized inner product produces high association between countries whose peaks in cases occurred at similar times. Indeed, the first cluster generally experienced much earlier peaks, as can be seen for Italy (\ref{fig:ItalyTS}) and Germany (\ref{fig:GermanyTS}), while the second cluster experienced large case counts much later, such as Brazil (\ref{fig:BrazilTS}), India (\ref{fig:IndiaTS}) and Argentina (\ref{fig:ArgentinaTS}). Even within the two subclusters, there is less similarity between case counts than there is for indices. This is reflected in the tree of Figure \ref{fig:equitycasesdend}, where branches belonging to countries' equity indices are split much lower in the tree's structure.

Most significantly, the figure reveals that there is no concurrence at all between large changes in countries' case counts and their equity indices. Excluding China's index, all other equity indices have moved together closely - even China itself exhibited some similarity with other indices.

The insights provided by Figure \ref{fig:equitycasesdend} are enhanced by the specific methodology presented above. For example, we trialed the commonly used Pearson correlation and distance correlation \cite{Szkely2007} measures between time series. The correlation does not perform as well due to the considerable flatness of the China cases time series, and does not identify it as an outlier. In addition, using correlation does not split the 16 countries (other than China) into two subclusters as described above. The distance correlation is not suitable in this setting: two time series $x_t$ and $y_t$ with any linear relationship, such as $x_t=ay_t + b$, are determined to have maximal distance correlation 1. This is not desired if $a$ is negative, meaning that $x_t$ increases or decreases in the opposite direction of $y_t$. As such, using distance correlation performs even more poorly, for instance allocating Canada and the UK to their own cluster and several other poor results.

\section{Conclusion}
\label{Conclusion}
In this paper, we analyze the natural history of COVID-19 across the world in conjunction with the stock market activity of 17 countries. Qualitatively and quantitatively we demonstrate that market movements have been highly uniform between these 17 countries, with China as the only exception.

In Section \ref{COVID19_case_death}, we analyze the structure of the multivariate time series of COVID-19 cases and deaths. Our analysis isolates the US as the unique member of the worst-affected cluster with respect to both cumulative cases and deaths for over a month, reflecting its exceptional impact by COVID-19. Subsequently, Brazil and India join that cluster, as their counts rose rapidly. The dendrograms in Figure \ref{datesdendrogram} each exhibit four contiguous intervals of dates, allowing us to observe key dates when the structure of the world's case counts changed substantially. With respect to cases, these dates are 05/30, 07/02, and 08/08. Indeed, all these dates herald significant shifts in the status of the disease around the world. On 05/30, Russia and Brazil enter the worst-affected cluster, on the same day as the latter reported a record number of cases \cite{Reutersbrazil}. On 07/02, several countries that had been heavily impacted earlier, such as China, South Korea, Singapore, and the Netherlands, enter less-affected clusters.  %further reduce clusters in severity.
This follows from June, a period of steady decline in Europe \cite{NPRJune}. On 08/08, both Singapore and the Netherlands move back into more severely affected clusters. Also around this time, India, Brazil, much of Africa and South America experience significantly more cases \cite{SMHAug}, while cases in Europe demonstrate a slower increase.

In Section \ref{International_trajectories}, we identify five characteristic behaviors of new case trajectories between countries. 24 countries exhibit their greatest counts up to smoothing on the final day of the period. 46 countries experience a second wave, with 28 experiencing a more severe second wave than the first. Singapore and Australia responded quickly to the virus \cite{australiabbc_2020}, and South Korea was hailed for its early contact tracing success \cite{Koreaguardian_2020}, yet all three of these countries experience second waves, and South Korea exhibits its greatest case counts at the end of the period. Italy and Spain were acknowledged to have imposed lockdowns too late in March \cite{italynyt2020}, with case counts eventually declining in May. Nonetheless, both of these countries experience second waves, with Spain's more severe than its first. Overall, long first waves and the return of second waves contribute to high case counts toward the end of the data period.

Despite the substantial activity in COVID-19 cases after March, the heterogeneity of subsequent waves and the number of countries with peaks in new cases, no discernible impact on financial indices was observed from March. In Section \ref{Equity_Market_Dynamics}, we apply a new method to analyze collective equity market dynamics across 17 countries in conjunction with their new case counts. Eigenvalue analysis indicates high similarity between 16 countries' equity prices, with China as the only outlier. We introduce an inner product pairing that demonstrates little concurrence between the profound market movements observed in March and development in COVID-19 cases.

Overall, we have chronicled the natural history of COVID-19 together with the market movements during 2020. Despite substantial heterogeneity in the new case trajectories on a country-by-country basis and frequent changes in the order and structure of most affected countries in cumulative cases, we have observed high homogeneity in the markets. All have moved together with substantial drawdown in March, followed by steady recovery, and no qualitative or quantitative relationship to any developments in COVID-19.

Our methods and results could be used by both financial practitioners and policymakers. The identification of a single country's index behaving differently from the rest of the collection, in this case China's, could provide opportunities for diversification of holdings and portfolio gains. More broadly, studying the sensitivity of financial securities to external factors is fundamental to market analysts and strategists. 
%This sensitivity may vary based on the asset class, size and stability of the security. 
The finding that the equity markets have been highly uniform, despite considerably differing trends in COVID-19 cases and deaths, may encourage policymakers to be more aggressive in pursuing containment measures. That is, the finding that the equity indices of countries that responded more aggressively to the pandemic have not been negatively impacted could provide confidence in prioritizing public health with reduced concern about harming their equity index.

Diverse future research possibilities exist with these methods. Market dynamics could be analyzed across different sectors, rather than countries. It is conceivable that during the pandemic, equity movements may be highly homogeneous within certain sectors such as technology, but heterogeneous when comparing retail and health care stocks. One could analyze fixed income assets such as government and corporate bonds, and determine if there is a bifurcation between equities and fixed income assets during market crises. Alternative asset classes such as cryptocurrencies could be studied, building on the work of \cite{Goodell2020}, who analyze the co-movement between COVID-19 and Bitcoin. One could also apply our analysis to alternative time series, such as %, such as alternative measures of volatility. For example, \cite{arouxet2020} use 
max-min and absolute return volatility \cite{arouxet2020,James2021_crypto}.
%when studying cryptocurrencies. 
Overall, the framework presented in this paper could be applied broadly to various economic, epidemiological and political crises.

%It could be of particular interest to study potential bifurcations in the impact market crises have among various asset classes (equities, fixed income, cryptocurrencies etc.) 

%----- END OF PAPER------------------

\section*{Data availability}
Daily COVID-19 case and death counts can be accessed at "Our World in Data" \cite{worldindata2020}. %Financial data is included in a supplementary file.

\section*{Acknowledgements}
Thanks to Kerry Chen for helpful edits and comments.

\appendix

\section{Cluster-based evolution methodology}
\label{appendix:clusterbased}
In this section, we provide more details and explanation for the methodology described in Section \ref{sec:clustermethodology}. Given the exponential spread of the disease, we select a logarithmic distance between counts. We replace any data entry that is empty or 0 - before any cases are detected - with a 1, so that the log of that entry is defined. Then, we define a distance on counts by $d(x,y)=|\log(x) - \log(y)|$. Effectively, this pulls back the Euclidean metric on $\mathbb{R}$ by the homeomorphism $\log: \mathbb{R}^+ \to \mathbb{R}$ and makes the positive reals a one-dimensional normed space. This allows us to use efficient cluster methods specific to one-dimensional data.

The goal is to partition the case or death counts $x_1(t),...,x_n(t)$ into a time-varying number of clusters at each time $t$. We wish to choose the number of clusters in such a way that provides us meaningful inference on how the multivariate time series evolves as a whole. A highly variable number of clusters would obscure inference on individual countries' cluster memberships. So we combine several methods of choosing this number to reduce any bias in our estimator and then implement exponential smoothing to yield a suitably changing number of clusters with time. For one-dimensional data, it is often regarded as unsuitable to use multivariate clustering methods, as simpler alternatives exist. We use an optimal implementation of K-means clustering called Ckmeans.1dp.dp \cite{Wang2011}. This requires the choice of the number of clusters \textit{a priori}.

To choose the number of clusters at each $t$, we average six methods described in \cite{Radchenko2017}. These methods are as follows: Ptbiserial index \cite{Milligan1980}, Silhouette score \cite{Rousseeuw1987}, KL index \cite{krzanowski1988}, C index \cite{Hubert1976}, McClain-Rao index \cite{Mcclain1975} and Dunn index \cite{Dunn1974}, but other methods could be used alternatively. Let the cluster numbers computed by these methods be $k_1(t),...,k_6(t)$, respectively. We define $k_{av}(t)=\frac{1}{6}\sum_{j=1}^6 k_j(t)$. This value is not necessarily an integer, so we cannot cluster with it directly. The function $k_{av}(t)$  is approximately locally stationary. So we may apply exponential smoothing to produce a smoothed integer value $\hat{k}(t).$ It is with this number that we cluster. Doing so at each $t$ produces a time-varying partition of the 92 countries into clusters, and defines an adjacency matrix $\text{Adj}^{(t)}$ for every time $t$.

\section{Turning point methodology}
\label{appendix:turningpoint}

In this section, we provide more details for the identification of turning points of a new case time series $z(t)$. First, some smoothing of the counts is necessary due to irregularities in the data set, and discrepancies between different data sources. There are consistently lower counts on the weekends, and some negative counts due to retroactive adjustments. The Savitzy-Golay filter ameliorates these issues by combining polynomial smoothing with a moving average computation - this moving average eliminates all but a few small negative counts; we simply replace these negative counts with zero. This yields a smoothed time series $\hat{z}(t) \in \mathbb{R}_{\geq 0}.$ Subsequently, we perform a two-step process to select and then refine a non-empty set $P$ of local maxima (peaks) and $T$ of local minima (troughs).

Following \cite{james2020covidusa}, we apply a two-step algorithm to the smoothed time series $\hat{z}(t)$. The first step produces an alternating sequence of troughs and peaks, beginning with a trough at $t=1$, where there are zero cases. The second step refines this sequence according to chosen conditions and parameters. The most important conditions to identify a peak or trough, respectively, in the first step, are the following:
\begin{align}
\label{baddefnpeak}
\hat{z}(t_0)&=\max\{\hat{z}(t): \max(1,t_0 - l) \leq t \leq \min(t_0 + l,T)\},\\
\label{baddefntrough}\hat{z}(t_0)&=\min\{\hat{z}(t): \max(1,t_0 - l) \leq t \leq \min(t_0 + l,T)\},
\end{align}
where $l$ is a parameter to be chosen. Following \cite{james2020covidusa}, we select $l=17$, which accounts for the 14-day incubation period of the virus \cite{incubation2020} and less testing on weekends. Defining peaks and troughs according to this definition alone has several flaws, such as the potential for two consecutive peaks.

Instead, we implement an inductive procedure to choose an alternating sequence of peaks and troughs. Suppose $t_0$ is the last determined peak. We search in the period $t>t_0$ for the first of two cases: if we find a time $t_1>t_0$ that satisfies (\ref{baddefntrough}) as well as a non-triviality condition $\hat{z}(t_1)<\hat{z}(t_0)$, we add $t_1$ to the set of troughs and proceed from there. If we find a time $t_1>t_0$ that satisfies (\ref{baddefnpeak}) and  $\hat{z}(t_0)\geq \hat{z}(t_1)$, we ignore this lower peak as redundant; if we find a time $t_1>t_0$ that satisfies (\ref{baddefnpeak}) and  $\hat{z}(t_1) > \hat{z}(t_0)$, we remove the peak $t_0$,  replace it with $t_1$ and continue from $t_1$. A similar process applies from a trough at $t_0$. 

At this point, the time series is assigned an alternating sequence of troughs and peaks. However, some turning points are immaterial and should be excluded. The second step is a flexible approach introduced in \cite{james2020covidusa} for this purpose. In this paper, we introduce new conditions within this framework. First, let $t_m$ be the global maximum of $\hat{z}(t)$. If this is not unique, we declare $t_m$ to be the first global maximum. This time $t_m$ is always declared a peak during the first step detailed above. Given any other peak $t_1$, we compute the peak ratio $\frac{\hat{z}(t_1)}{\hat{z}(t_m)}$. We select a parameter $\delta$, and if $\frac{\hat{z}(t_1)}{\hat{z}(t_m)}<\delta$, we remove the peak $t_1$. If two consecutive troughs $t_0,t_2$ remain, we remove $t_0$ if $\hat{z}(t_0)>\hat{z}(t_2)$, and remove $t_2$ if $\hat{z}(t_0)\leq\hat{z}(t_2)$. That is, we ensure the sequence of peaks and troughs remains alternating. In our implementation, we choose $\delta=0.05.$ Unlike \cite{james2020covidusa}, we remove earlier peaks, not just subsequent peaks, according to this condition.

Finally, we use the same \emph{log-gradient} function between times $t_1<t_2$, defined as
\begin{align}
\label{loggrad}
   \loggrad(t_1,t_2)=\frac{\log \hat{z}(t_2) - \log \hat{z}(t_1)}{t_2-t_1}.
\end{align}
The numerator equals  $\log(\frac{\hat{z}(t_2)}{\hat{z}(t_1)})$, a "logarithmic rate of change." Unlike the standard rate of change given by $\frac{\hat{z}(t_2)}{\hat{z}(t_1)} -1$, the logarithmic change is symmetrically between $(-\infty,\infty)$. Let $t_1,t_2$ be adjacent turning points (one a trough, one a peak). We choose a parameter $\epsilon=0.007$;  if
\begin{align}
    |\loggrad(t_1,t_2)|<\epsilon,
\end{align}
that is, the average logarithmic change is less than 0.7\%, we remove $t_2$ from our sets of peaks and troughs. If $t_2$ is not the final turning point, we also remove $t_1$.

% Table for phenomena
\section{Classification of countries by wave behavior}
\label{sec:classification}
In Table \ref{tab:behavior}, we classify all $n=92$ countries into 5 different characteristic classes according to the methodology of Section \ref{International_trajectories}.

\begin{table}
\begin{tabular}{|p{1.8cm}|p{3.2cm}|p{3.3cm}|p{2.8cm}|p{2cm}|}
 \hline
 \multicolumn{5}{|c|}{Country behaviors} \\
 \hline
 First wave & Over first wave & Second wave & Over second wave & Third wave \\
 \hline
  \{TP\} &  \{TPT\} & \{TPTP\} & \{TPTPT\} & \{TPTPTP\} \\
 \hline
 Argentina & Afghanistan & Austria & Algeria & Croatia \\
 Brazil & Armenia & Belarus & Australia & Greece \\
 Colombia & Azerbaijan & Bosnia \& Herzegovina & Belgium & Portugal \\
 Costa Rica & Bahrain & Canada & Bulgaria & South Korea \\
 India & Bangladesh & Czech Republic & Israel &  \\
 Indonesia & Bolivia & Denmark & Japan &  \\
 Iraq & Cameroon & Ecuador & Kosovo &  \\
 Lebanon & Chile & Ethiopia & North Macedonia &  \\
 Libya & China & France & Panama &  \\
 Palestine & C{\^o}te d'Ivoire & Germany & Serbia &  \\
 Paraguay & Dominican Republic & Honduras & Singapore &  \\
 Philippines & DR Congo & Iran & United States &  \\
 Senegal & Egypt & Ireland & Uzbekistan &  \\
 Ukraine & El Salvador & Italy & Zambia &  \\
 Venezuela & Ghana & Moldova &   &  \\
  & Guatemala & Morocco &   &  \\
  & Kazakhstan & Nepal &   &  \\
  & Kenya & Netherlands &   &  \\
  & Kuwait & Norway &   &  \\
  & Kyrgyzstan & Peru &   &  \\
  & Madagascar & Poland &   &  \\
  & Mexico & Romania &   &  \\
  & Nigeria & Spain &   &  \\
  & Oman & Sudan &   &  \\
 & Pakistan & Switzerland &  &  \\
  & Puerto Rico & Turkey &   &  \\
  & Qatar & United Arab Emirates &  &  \\
  & Russia & United Kingdom &   &  \\
  & Saudi Arabia &   &   &  \\
  & South Africa &   &  &  \\
  & Sweden &  &   &  \\
\hline
\end{tabular}
\caption{Classification of 92 countries according to the methodology of Section \ref{International_trajectories}. There are 15 countries in their first wave, 31 countries in a decline from or completely over their first wave, 28 countries in their second wave, 14 countries in a decline from or completely over their second wave and 4 countries in their third wave.}
\label{tab:behavior}
\end{table}

\clearpage
\bibliographystyle{elsarticle-num-names}
\bibliography{References.bib}

\begin{thebibliography}{88}
\expandafter\ifx\csname natexlab\endcsname\relax\def\natexlab#1{#1}\fi
\providecommand{\url}[1]{\texttt{#1}}
\providecommand{\href}[2]{#2}
\providecommand{\path}[1]{#1}
\providecommand{\DOIprefix}{doi:}
\providecommand{\ArXivprefix}{arXiv:}
\providecommand{\URLprefix}{URL: }
\providecommand{\Pubmedprefix}{pmid:}
\providecommand{\doi}[1]{\href{http://dx.doi.org/#1}{\path{#1}}}
\providecommand{\Pubmed}[1]{\href{pmid:#1}{\path{#1}}}
\providecommand{\bibinfo}[2]{#2}
\ifx\xfnm\relax \def\xfnm[#1]{\unskip,\space#1}\fi
%Type = Article
\bibitem[{Wang et~al.(2020)Wang, Zhang, Zhao, Zhang, and Jiang}]{Wang2020}
\bibinfo{author}{G.~Wang}, \bibinfo{author}{Y.~Zhang},
  \bibinfo{author}{J.~Zhao}, \bibinfo{author}{J.~Zhang},
  \bibinfo{author}{F.~Jiang},
\newblock \bibinfo{title}{Mitigate the effects of home confinement on children
  during the {COVID}-19 outbreak},
\newblock \bibinfo{journal}{The Lancet} \bibinfo{volume}{395}
  (\bibinfo{year}{2020}) \bibinfo{pages}{945--947}.
  \DOIprefix\doi{10.1016/s0140-6736(20)30547-x}.
%Type = Article
\bibitem[{Chinazzi et~al.(2020)}]{Chinazzi2020}
\bibinfo{author}{M.~Chinazzi}, et~al.,
\newblock \bibinfo{title}{The effect of travel restrictions on the spread of
  the 2019 novel coronavirus ({COVID}-19) outbreak},
\newblock \bibinfo{journal}{Science} \bibinfo{volume}{368}
  (\bibinfo{year}{2020}) \bibinfo{pages}{395--400}.
  \DOIprefix\doi{10.1126/science.aba9757}.
%Type = Article
\bibitem[{Liu et~al.(2020)Liu, Gayle, Wilder-Smith, and Rockl\"{o}v}]{Liu2020}
\bibinfo{author}{Y.~Liu}, \bibinfo{author}{A.~A. Gayle},
  \bibinfo{author}{A.~Wilder-Smith}, \bibinfo{author}{J.~Rockl\"{o}v},
\newblock \bibinfo{title}{The reproductive number of {COVID}-19 is higher
  compared to {SARS} coronavirus},
\newblock \bibinfo{journal}{Journal of Travel Medicine} \bibinfo{volume}{27}
  (\bibinfo{year}{2020}). \DOIprefix\doi{10.1093/jtm/taaa021}.
%Type = Article
\bibitem[{Fang et~al.(2020)Fang, Nie, and Penny}]{Fang2020}
\bibinfo{author}{Y.~Fang}, \bibinfo{author}{Y.~Nie},
  \bibinfo{author}{M.~Penny},
\newblock \bibinfo{title}{Transmission dynamics of the {COVID}-19 outbreak and
  effectiveness of government interventions: A data-driven analysis},
\newblock \bibinfo{journal}{Journal of Medical Virology} \bibinfo{volume}{92}
  (\bibinfo{year}{2020}) \bibinfo{pages}{645--659}.
  \DOIprefix\doi{10.1002/jmv.25750}.
%Type = Article
\bibitem[{Zhou et~al.(2020)}]{Zhou2020}
\bibinfo{author}{P.~Zhou}, et~al.,
\newblock \bibinfo{title}{A pneumonia outbreak associated with a new
  coronavirus of probable bat origin},
\newblock \bibinfo{journal}{Nature} \bibinfo{volume}{579}
  (\bibinfo{year}{2020}) \bibinfo{pages}{270--273}.
  \DOIprefix\doi{10.1038/s41586-020-2012-7}.
%Type = Article
\bibitem[{Dehning et~al.(2020)Dehning, Zierenberg, Spitzner, Wibral, Neto,
  Wilczek, and Priesemann}]{Dehning2020}
\bibinfo{author}{J.~Dehning}, \bibinfo{author}{J.~Zierenberg},
  \bibinfo{author}{F.~P. Spitzner}, \bibinfo{author}{M.~Wibral},
  \bibinfo{author}{J.~P. Neto}, \bibinfo{author}{M.~Wilczek},
  \bibinfo{author}{V.~Priesemann},
\newblock \bibinfo{title}{Inferring change points in the spread of {COVID}-19
  reveals the effectiveness of interventions},
\newblock \bibinfo{journal}{Science} \bibinfo{volume}{369}
  (\bibinfo{year}{2020}) \bibinfo{pages}{eabb9789}.
  \DOIprefix\doi{10.1126/science.abb9789}.
%Type = Article
\bibitem[{Jiang et~al.(2020)Jiang, Deng, Zhang, Cai, Cheung, and
  Xia}]{Jiang2020}
\bibinfo{author}{F.~Jiang}, \bibinfo{author}{L.~Deng},
  \bibinfo{author}{L.~Zhang}, \bibinfo{author}{Y.~Cai}, \bibinfo{author}{C.~W.
  Cheung}, \bibinfo{author}{Z.~Xia},
\newblock \bibinfo{title}{Review of the clinical characteristics of
  {C}oronavirus {D}isease 2019 ({COVID}-19)},
\newblock \bibinfo{journal}{Journal of General Internal Medicine}
  \bibinfo{volume}{35} (\bibinfo{year}{2020}) \bibinfo{pages}{1545--1549}.
  \DOIprefix\doi{10.1007/s11606-020-05762-w}.
%Type = Article
\bibitem[{Zu et~al.(2020)Zu, Jiang, Xu, Chen, Ni, Lu, and Zhang}]{Zu2020}
\bibinfo{author}{Z.~Y. Zu}, \bibinfo{author}{M.~D. Jiang},
  \bibinfo{author}{P.~P. Xu}, \bibinfo{author}{W.~Chen}, \bibinfo{author}{Q.~Q.
  Ni}, \bibinfo{author}{G.~M. Lu}, \bibinfo{author}{L.~J. Zhang},
\newblock \bibinfo{title}{Coronavirus disease 2019 ({COVID}-19): A perspective
  from china},
\newblock \bibinfo{journal}{Radiology}  (\bibinfo{year}{2020})
  \bibinfo{pages}{200490}. \DOIprefix\doi{10.1148/radiol.2020200490}.
%Type = Article
\bibitem[{Li and Clercq(2020)}]{Li2020}
\bibinfo{author}{G.~Li}, \bibinfo{author}{E.~D. Clercq},
\newblock \bibinfo{title}{Therapeutic options for the 2019 novel coronavirus
  (2019-{nCoV})},
\newblock \bibinfo{journal}{Nature Reviews Drug Discovery} \bibinfo{volume}{19}
  (\bibinfo{year}{2020}) \bibinfo{pages}{149--150}.
  \DOIprefix\doi{10.1038/d41573-020-00016-0}.
%Type = Article
\bibitem[{Zhang and Liu(2020)}]{Zhang2020}
\bibinfo{author}{L.~Zhang}, \bibinfo{author}{Y.~Liu},
\newblock \bibinfo{title}{Potential interventions for novel coronavirus in
  {C}hina: A systematic review},
\newblock \bibinfo{journal}{Journal of Medical Virology} \bibinfo{volume}{92}
  (\bibinfo{year}{2020}) \bibinfo{pages}{479--490}.
  \DOIprefix\doi{10.1002/jmv.25707}.
%Type = Article
\bibitem[{Wang et~al.(2020)Wang, Cao, Zhang, Yang, Liu, Xu, Shi, Hu, Zhong, and
  Xiao}]{Wang2020_CELL}
\bibinfo{author}{M.~Wang}, \bibinfo{author}{R.~Cao},
  \bibinfo{author}{L.~Zhang}, \bibinfo{author}{X.~Yang},
  \bibinfo{author}{J.~Liu}, \bibinfo{author}{M.~Xu}, \bibinfo{author}{Z.~Shi},
  \bibinfo{author}{Z.~Hu}, \bibinfo{author}{W.~Zhong},
  \bibinfo{author}{G.~Xiao},
\newblock \bibinfo{title}{Remdesivir and chloroquine effectively inhibit the
  recently emerged novel coronavirus (2019-{nCoV}) in vitro},
\newblock \bibinfo{journal}{Cell Research} \bibinfo{volume}{30}
  (\bibinfo{year}{2020}) \bibinfo{pages}{269--271}.
  \DOIprefix\doi{10.1038/s41422-020-0282-0}.
%Type = Article
\bibitem[{Cao et~al.(2020)}]{NEJM2020}
\bibinfo{author}{B.~Cao}, et~al.,
\newblock \bibinfo{title}{A trial of {Lopinavir-Ritonavir} in {C}ovid-19},
\newblock \bibinfo{journal}{New England Journal of Medicine}
  \bibinfo{volume}{382} (\bibinfo{year}{2020}) \bibinfo{pages}{e68}.
  \DOIprefix\doi{10.1056/nejmc2008043}.
%Type = Article
\bibitem[{Corey et~al.(2020)Corey, Mascola, Fauci, and Collins}]{Corey2020}
\bibinfo{author}{L.~Corey}, \bibinfo{author}{J.~R. Mascola},
  \bibinfo{author}{A.~S. Fauci}, \bibinfo{author}{F.~S. Collins},
\newblock \bibinfo{title}{A strategic approach to {COVID}-19 vaccine
  {R}{\&}{D}},
\newblock \bibinfo{journal}{Science} \bibinfo{volume}{368}
  (\bibinfo{year}{2020}) \bibinfo{pages}{948--950}.
  \DOIprefix\doi{10.1126/science.abc5312}.
%Type = Article
\bibitem[{Zhang et~al.(2020)Zhang, Hu, and Ji}]{Zhang2020finance}
\bibinfo{author}{D.~Zhang}, \bibinfo{author}{M.~Hu}, \bibinfo{author}{Q.~Ji},
\newblock \bibinfo{title}{Financial markets under the global pandemic of
  {COVID}-19},
\newblock \bibinfo{journal}{Finance Research Letters}  (\bibinfo{year}{2020})
  \bibinfo{pages}{101528}. \DOIprefix\doi{10.1016/j.frl.2020.101528}.
%Type = Article
\bibitem[{He et~al.(2020)He, Liu, Wang, and Yu}]{He2020}
\bibinfo{author}{Q.~He}, \bibinfo{author}{J.~Liu}, \bibinfo{author}{S.~Wang},
  \bibinfo{author}{J.~Yu},
\newblock \bibinfo{title}{The impact of {COVID}-19 on stock markets},
\newblock \bibinfo{journal}{Economic and Political Studies}
  (\bibinfo{year}{2020}) \bibinfo{pages}{1--14}.
  \DOIprefix\doi{10.1080/20954816.2020.1757570}.
%Type = Article
\bibitem[{Zaremba et~al.(2020)Zaremba, Kizys, Aharon, and Demir}]{Zaremba2020}
\bibinfo{author}{A.~Zaremba}, \bibinfo{author}{R.~Kizys},
  \bibinfo{author}{D.~Y. Aharon}, \bibinfo{author}{E.~Demir},
\newblock \bibinfo{title}{Infected markets: Novel coronavirus, government
  interventions, and stock return volatility around the globe},
\newblock \bibinfo{journal}{Finance Research Letters} \bibinfo{volume}{35}
  (\bibinfo{year}{2020}) \bibinfo{pages}{101597}.
  \DOIprefix\doi{10.1016/j.frl.2020.101597}.
%Type = Article
\bibitem[{Akhtaruzzaman et~al.(2020)Akhtaruzzaman, Boubaker, and
  Sensoy}]{Akhtaruzzaman2020}
\bibinfo{author}{M.~Akhtaruzzaman}, \bibinfo{author}{S.~Boubaker},
  \bibinfo{author}{A.~Sensoy},
\newblock \bibinfo{title}{Financial contagion during {COVID}{\textendash}19
  crisis},
\newblock \bibinfo{journal}{Finance Research Letters}  (\bibinfo{year}{2020})
  \bibinfo{pages}{101604}. \URLprefix
  \url{https://doi.org/10.1016/j.frl.2020.101604}.
  \DOIprefix\doi{10.1016/j.frl.2020.101604}.
%Type = Article
\bibitem[{Okorie and Lin(2020)}]{Okorie2020}
\bibinfo{author}{D.~I. Okorie}, \bibinfo{author}{B.~Lin},
\newblock \bibinfo{title}{Stock markets and the {COVID}-19 fractal contagion
  effects},
\newblock \bibinfo{journal}{Finance Research Letters}  (\bibinfo{year}{2020})
  \bibinfo{pages}{101640}. \DOIprefix\doi{10.1016/j.frl.2020.101640}.
%Type = Article
\bibitem[{Lahmiri and Bekiros(2020)}]{Lahmiri2020}
\bibinfo{author}{S.~Lahmiri}, \bibinfo{author}{S.~Bekiros},
\newblock \bibinfo{title}{The impact of {COVID}-19 pandemic upon stability and
  sequential irregularity of equity and cryptocurrency markets},
\newblock \bibinfo{journal}{Chaos, Solitons {\&} Fractals}
  \bibinfo{volume}{138} (\bibinfo{year}{2020}) \bibinfo{pages}{109936}.
  \DOIprefix\doi{10.1016/j.chaos.2020.109936}.
%Type = Article
\bibitem[{Khajanchi and Sarkar(2020)}]{Khajanchi2020}
\bibinfo{author}{S.~Khajanchi}, \bibinfo{author}{K.~Sarkar},
\newblock \bibinfo{title}{Forecasting the daily and cumulative number of cases
  for the {COVID}-19 pandemic in {India}},
\newblock \bibinfo{journal}{Chaos: An Interdisciplinary Journal of Nonlinear
  Science} \bibinfo{volume}{30} (\bibinfo{year}{2020}) \bibinfo{pages}{071101}.
  \DOIprefix\doi{10.1063/5.0016240}.
%Type = Article
\bibitem[{Ribeiro et~al.(2020)Ribeiro, da~Silva, Mariani, and dos
  Santos~Coelho}]{Ribeiro2020}
\bibinfo{author}{M.~H. D.~M. Ribeiro}, \bibinfo{author}{R.~G. da~Silva},
  \bibinfo{author}{V.~C. Mariani}, \bibinfo{author}{L.~dos Santos~Coelho},
\newblock \bibinfo{title}{Short-term forecasting {COVID}-19 cumulative
  confirmed cases: Perspectives for {B}razil},
\newblock \bibinfo{journal}{Chaos, Solitons {\&} Fractals}
  \bibinfo{volume}{135} (\bibinfo{year}{2020}) \bibinfo{pages}{109853}.
  \DOIprefix\doi{10.1016/j.chaos.2020.109853}.
%Type = Article
\bibitem[{Chakraborty and Ghosh(2020)}]{Chakraborty2020}
\bibinfo{author}{T.~Chakraborty}, \bibinfo{author}{I.~Ghosh},
\newblock \bibinfo{title}{Real-time forecasts and risk assessment of novel
  coronavirus ({COVID}-19) cases: A data-driven analysis},
\newblock \bibinfo{journal}{Chaos, Solitons {\&} Fractals}
  \bibinfo{volume}{135} (\bibinfo{year}{2020}) \bibinfo{pages}{109850}.
  \DOIprefix\doi{10.1016/j.chaos.2020.109850}.
%Type = Article
\bibitem[{James and Menzies(2020)}]{James2020_chaos}
\bibinfo{author}{N.~James}, \bibinfo{author}{M.~Menzies},
\newblock \bibinfo{title}{Cluster-based dual evolution for multivariate time
  series: Analyzing {COVID}-19},
\newblock \bibinfo{journal}{Chaos: An Interdisciplinary Journal of Nonlinear
  Science} \bibinfo{volume}{30} (\bibinfo{year}{2020}) \bibinfo{pages}{061108}.
  \DOIprefix\doi{10.1063/5.0013156}.
%Type = Article
\bibitem[{Ballesteros et~al.(2020)Ballesteros, Blasco, and
  Gutierrez-Sagredo}]{SIRBallesteros2020}
\bibinfo{author}{A.~Ballesteros}, \bibinfo{author}{A.~Blasco},
  \bibinfo{author}{I.~Gutierrez-Sagredo},
\newblock \bibinfo{title}{Hamiltonian structure of compartmental
  epidemiological models},
\newblock \bibinfo{journal}{Physica D: Nonlinear Phenomena}
  \bibinfo{volume}{413} (\bibinfo{year}{2020}) \bibinfo{pages}{132656}.
  \DOIprefix\doi{10.1016/j.physd.2020.132656}.
%Type = Article
\bibitem[{Barlow and Weinstein(2020)}]{SIRBarlow2020}
\bibinfo{author}{N.~S. Barlow}, \bibinfo{author}{S.~J. Weinstein},
\newblock \bibinfo{title}{Accurate closed-form solution of the {SIR} epidemic
  model},
\newblock \bibinfo{journal}{Physica D: Nonlinear Phenomena}
  \bibinfo{volume}{408} (\bibinfo{year}{2020}) \bibinfo{pages}{132540}.
  \DOIprefix\doi{10.1016/j.physd.2020.132540}.
%Type = Article
\bibitem[{Cadoni and Gaeta(2020)}]{SIRCadoni2020}
\bibinfo{author}{M.~Cadoni}, \bibinfo{author}{G.~Gaeta},
\newblock \bibinfo{title}{Size and timescale of epidemics in the {SIR}
  framework},
\newblock \bibinfo{journal}{Physica D: Nonlinear Phenomena}
  \bibinfo{volume}{411} (\bibinfo{year}{2020}) \bibinfo{pages}{132626}.
  \DOIprefix\doi{10.1016/j.physd.2020.132626}.
%Type = Article
\bibitem[{Comunian et~al.(2020)Comunian, Gaburro, and
  Giudici}]{SIRComunian2020}
\bibinfo{author}{A.~Comunian}, \bibinfo{author}{R.~Gaburro},
  \bibinfo{author}{M.~Giudici},
\newblock \bibinfo{title}{Inversion of a {SIR}-based model: A critical analysis
  about the application to {COVID}-19 epidemic},
\newblock \bibinfo{journal}{Physica D: Nonlinear Phenomena}
  \bibinfo{volume}{413} (\bibinfo{year}{2020}) \bibinfo{pages}{132674}.
  \DOIprefix\doi{10.1016/j.physd.2020.132674}.
%Type = Article
\bibitem[{Neves and Guerrero(2020)}]{SIRNeves2020}
\bibinfo{author}{A.~G. Neves}, \bibinfo{author}{G.~Guerrero},
\newblock \bibinfo{title}{Predicting the evolution of the {COVID}-19 epidemic
  with the a-{SIR} model: Lombardy, {I}taly and {S}{\~{a}}o {P}aulo state,
  {B}razil},
\newblock \bibinfo{journal}{Physica D: Nonlinear Phenomena}
  \bibinfo{volume}{413} (\bibinfo{year}{2020}) \bibinfo{pages}{132693}.
  \DOIprefix\doi{10.1016/j.physd.2020.132693}.
%Type = Article
\bibitem[{Vyasarayani and Chatterjee(2020)}]{SIRVyasarayani2020}
\bibinfo{author}{C.~Vyasarayani}, \bibinfo{author}{A.~Chatterjee},
\newblock \bibinfo{title}{New approximations, and policy implications, from a
  delayed dynamic model of a fast pandemic},
\newblock \bibinfo{journal}{Physica D: Nonlinear Phenomena}
  \bibinfo{volume}{414} (\bibinfo{year}{2020}) \bibinfo{pages}{132701}.
  \DOIprefix\doi{10.1016/j.physd.2020.132701}.
%Type = Article
\bibitem[{Weinstein et~al.(2020)Weinstein, Holland, Rogers, and
  Barlow}]{SIRWeinstein2020}
\bibinfo{author}{S.~J. Weinstein}, \bibinfo{author}{M.~S. Holland},
  \bibinfo{author}{K.~E. Rogers}, \bibinfo{author}{N.~S. Barlow},
\newblock \bibinfo{title}{Analytic solution of the {SEIR} epidemic model via
  asymptotic approximant},
\newblock \bibinfo{journal}{Physica D: Nonlinear Phenomena}
  \bibinfo{volume}{411} (\bibinfo{year}{2020}) \bibinfo{pages}{132633}.
  \DOIprefix\doi{10.1016/j.physd.2020.132633}.
%Type = Article
\bibitem[{Beare and Toda(2020)}]{Beare2020}
\bibinfo{author}{B.~K. Beare}, \bibinfo{author}{A.~A. Toda},
\newblock \bibinfo{title}{On the emergence of a power law in the distribution
  of {COVID}-19 cases},
\newblock \bibinfo{journal}{Physica D: Nonlinear Phenomena}
  \bibinfo{volume}{412} (\bibinfo{year}{2020}) \bibinfo{pages}{132649}.
  \DOIprefix\doi{10.1016/j.physd.2020.132649}.
%Type = Article
\bibitem[{Manchein et~al.(2020)Manchein, Brugnago, da~Silva, Mendes, and
  Beims}]{Manchein2020}
\bibinfo{author}{C.~Manchein}, \bibinfo{author}{E.~L. Brugnago},
  \bibinfo{author}{R.~M. da~Silva}, \bibinfo{author}{C.~F.~O. Mendes},
  \bibinfo{author}{M.~W. Beims},
\newblock \bibinfo{title}{Strong correlations between power-law growth of
  {COVID}-19 in four continents and the inefficiency of soft quarantine
  strategies},
\newblock \bibinfo{journal}{Chaos: An Interdisciplinary Journal of Nonlinear
  Science} \bibinfo{volume}{30} (\bibinfo{year}{2020}) \bibinfo{pages}{041102}.
  \DOIprefix\doi{10.1063/5.0009454}.
%Type = Article
\bibitem[{Blasius(2020)}]{Blasius2020}
\bibinfo{author}{B.~Blasius},
\newblock \bibinfo{title}{Power-law distribution in the number of confirmed
  {COVID}-19 cases},
\newblock \bibinfo{journal}{Chaos: An Interdisciplinary Journal of Nonlinear
  Science} \bibinfo{volume}{30} (\bibinfo{year}{2020}) \bibinfo{pages}{093123}.
  \DOIprefix\doi{10.1063/5.0013031}.
%Type = Article
\bibitem[{Anastassopoulou et~al.(2020)Anastassopoulou, Russo, Tsakris, and
  Siettos}]{Anastassopoulou2020}
\bibinfo{author}{C.~Anastassopoulou}, \bibinfo{author}{L.~Russo},
  \bibinfo{author}{A.~Tsakris}, \bibinfo{author}{C.~Siettos},
\newblock \bibinfo{title}{Data-based analysis, modelling and forecasting of the
  {COVID}-19 outbreak},
\newblock \bibinfo{journal}{{PLOS} {ONE}} \bibinfo{volume}{15}
  (\bibinfo{year}{2020}) \bibinfo{pages}{e0230405}.
  \DOIprefix\doi{10.1371/journal.pone.0230405}.
%Type = Article
\bibitem[{Vazquez(2006)}]{Vazquez2006}
\bibinfo{author}{A.~Vazquez},
\newblock \bibinfo{title}{Polynomial growth in branching processes with
  diverging reproductive number},
\newblock \bibinfo{journal}{Physical Review Letters} \bibinfo{volume}{96}
  (\bibinfo{year}{2006}). \DOIprefix\doi{10.1103/physrevlett.96.038702}.
%Type = Article
\bibitem[{Gopikrishnan et~al.(1998)Gopikrishnan, Meyer, Amaral, and
  Stanley}]{Gopikrishnan1998}
\bibinfo{author}{P.~Gopikrishnan}, \bibinfo{author}{M.~Meyer},
  \bibinfo{author}{L.~Amaral}, \bibinfo{author}{H.~Stanley},
\newblock \bibinfo{title}{Inverse cubic law for the distribution of stock price
  variations},
\newblock \bibinfo{journal}{The European Physical Journal B}
  \bibinfo{volume}{3} (\bibinfo{year}{1998}) \bibinfo{pages}{139--140}.
  \DOIprefix\doi{10.1007/s100510050292}.
%Type = Article
\bibitem[{Podobnik et~al.(2009)Podobnik, Horvatic, Petersen, and
  Stanley}]{Podobnik2009}
\bibinfo{author}{B.~Podobnik}, \bibinfo{author}{D.~Horvatic},
  \bibinfo{author}{A.~M. Petersen}, \bibinfo{author}{H.~E. Stanley},
\newblock \bibinfo{title}{Cross-correlations between volume change and price
  change},
\newblock \bibinfo{journal}{Proceedings of the National Academy of Sciences}
  \bibinfo{volume}{106} (\bibinfo{year}{2009}) \bibinfo{pages}{22079--22084}.
  \DOIprefix\doi{10.1073/pnas.0911983106}.
%Type = Article
\bibitem[{Liu et~al.(1999)Liu, Gopikrishnan, Cizeau, Meyer, Peng, and
  Stanley}]{Liu1999}
\bibinfo{author}{Y.~Liu}, \bibinfo{author}{P.~Gopikrishnan},
  \bibinfo{author}{Cizeau}, \bibinfo{author}{Meyer}, \bibinfo{author}{Peng},
  \bibinfo{author}{H.~E. Stanley},
\newblock \bibinfo{title}{Statistical properties of the volatility of price
  fluctuations},
\newblock \bibinfo{journal}{Physical Review E} \bibinfo{volume}{60}
  (\bibinfo{year}{1999}) \bibinfo{pages}{1390--1400}.
  \DOIprefix\doi{10.1103/physreve.60.1390}.
%Type = Article
\bibitem[{Moeckel and Murray(1997)}]{Moeckel1997}
\bibinfo{author}{R.~Moeckel}, \bibinfo{author}{B.~Murray},
\newblock \bibinfo{title}{Measuring the distance between time series},
\newblock \bibinfo{journal}{Physica D: Nonlinear Phenomena}
  \bibinfo{volume}{102} (\bibinfo{year}{1997}) \bibinfo{pages}{187--194}.
  \DOIprefix\doi{10.1016/s0167-2789(96)00154-6}.
%Type = Article
\bibitem[{Sz{\'{e}}kely et~al.(2007)Sz{\'{e}}kely, Rizzo, and
  Bakirov}]{Szkely2007}
\bibinfo{author}{G.~J. Sz{\'{e}}kely}, \bibinfo{author}{M.~L. Rizzo},
  \bibinfo{author}{N.~K. Bakirov},
\newblock \bibinfo{title}{Measuring and testing dependence by correlation of
  distances},
\newblock \bibinfo{journal}{The Annals of Statistics} \bibinfo{volume}{35}
  (\bibinfo{year}{2007}) \bibinfo{pages}{2769--2794}.
  \DOIprefix\doi{10.1214/009053607000000505}.
%Type = Article
\bibitem[{Mendes and Beims(2018)}]{Mendes2018}
\bibinfo{author}{C.~F. Mendes}, \bibinfo{author}{M.~W. Beims},
\newblock \bibinfo{title}{Distance correlation detecting {L}yapunov
  instabilities, noise-induced escape times and mixing},
\newblock \bibinfo{journal}{Physica A: Statistical Mechanics and its
  Applications} \bibinfo{volume}{512} (\bibinfo{year}{2018})
  \bibinfo{pages}{721--730}. \DOIprefix\doi{10.1016/j.physa.2018.08.028}.
%Type = Article
\bibitem[{Mendes et~al.(2019)Mendes, da~Silva, and Beims}]{Mendes2019}
\bibinfo{author}{C.~F.~O. Mendes}, \bibinfo{author}{R.~M. da~Silva},
  \bibinfo{author}{M.~W. Beims},
\newblock \bibinfo{title}{Decay of the distance autocorrelation and {L}yapunov
  exponents},
\newblock \bibinfo{journal}{Physical Review E} \bibinfo{volume}{99}
  (\bibinfo{year}{2019}). \DOIprefix\doi{10.1103/physreve.99.062206}.
%Type = Article
\bibitem[{Shang et~al.(2020)Shang, Yang, Moore, Ji, and Small}]{Shang2020}
\bibinfo{author}{K.~Shang}, \bibinfo{author}{B.~Yang}, \bibinfo{author}{J.~M.
  Moore}, \bibinfo{author}{Q.~Ji}, \bibinfo{author}{M.~Small},
\newblock \bibinfo{title}{Growing networks with communities: A distributive
  link model},
\newblock \bibinfo{journal}{Chaos: An Interdisciplinary Journal of Nonlinear
  Science} \bibinfo{volume}{30} (\bibinfo{year}{2020}) \bibinfo{pages}{041101}.
  \DOIprefix\doi{10.1063/5.0007422}.
%Type = Article
\bibitem[{Onnela et~al.(2004)Onnela, Kaski, and Kert{'e}sz}]{Onnela2004}
\bibinfo{author}{J.-P. Onnela}, \bibinfo{author}{K.~Kaski},
  \bibinfo{author}{J.~Kert{'e}sz},
\newblock \bibinfo{title}{Clustering and information in correlation based
  financial networks},
\newblock \bibinfo{journal}{The European Physical Journal B - Condensed Matter}
  \bibinfo{volume}{38} (\bibinfo{year}{2004}) \bibinfo{pages}{353--362}.
  \DOIprefix\doi{10.1140/epjb/e2004-00128-7}.
%Type = Article
\bibitem[{Fenn et~al.(2011)Fenn, Porter, Williams, McDonald, Johnson, and
  Jones}]{Fenn2011}
\bibinfo{author}{D.~J. Fenn}, \bibinfo{author}{M.~A. Porter},
  \bibinfo{author}{S.~Williams}, \bibinfo{author}{M.~McDonald},
  \bibinfo{author}{N.~F. Johnson}, \bibinfo{author}{N.~S. Jones},
\newblock \bibinfo{title}{Temporal evolution of financial-market correlations},
\newblock \bibinfo{journal}{Physical Review E} \bibinfo{volume}{84}
  (\bibinfo{year}{2011}). \URLprefix
  \url{https://doi.org/10.1103/physreve.84.026109}.
  \DOIprefix\doi{10.1103/physreve.84.026109}.
%Type = Article
\bibitem[{Drozdz et~al.(2018)Drozdz, Gebarowski, Minati, Oswiecimka, and
  Wactorek}]{Drod2018}
\bibinfo{author}{S.~Drozdz}, \bibinfo{author}{R.~Gebarowski},
  \bibinfo{author}{L.~Minati}, \bibinfo{author}{P.~Oswiecimka},
  \bibinfo{author}{M.~Wactorek},
\newblock \bibinfo{title}{Bitcoin market route to maturity? {E}vidence from
  return fluctuations, temporal correlations and multiscaling effects},
\newblock \bibinfo{journal}{Chaos: An Interdisciplinary Journal of Nonlinear
  Science} \bibinfo{volume}{28} (\bibinfo{year}{2018}) \bibinfo{pages}{071101}.
  \URLprefix \url{https://doi.org/10.1063/1.5036517}.
  \DOIprefix\doi{10.1063/1.5036517}.
%Type = Article
\bibitem[{Drozdz et~al.(2020)Drozdz, Minati, Oswiecimka, Stanuszek, and
  Wactorek}]{Drod2020}
\bibinfo{author}{S.~Drozdz}, \bibinfo{author}{L.~Minati},
  \bibinfo{author}{P.~Oswiecimka}, \bibinfo{author}{M.~Stanuszek},
  \bibinfo{author}{M.~Wactorek},
\newblock \bibinfo{title}{Competition of noise and collectivity in global
  cryptocurrency trading: Route to a self-contained market},
\newblock \bibinfo{journal}{Chaos: An Interdisciplinary Journal of Nonlinear
  Science} \bibinfo{volume}{30} (\bibinfo{year}{2020}) \bibinfo{pages}{023122}.
  \DOIprefix\doi{10.1063/1.5139634}.
%Type = Article
\bibitem[{Eisler and Kert{\'{e}}sz(2006)}]{Eisler2006}
\bibinfo{author}{Z.~Eisler}, \bibinfo{author}{J.~Kert{\'{e}}sz},
\newblock \bibinfo{title}{Scaling theory of temporal correlations and
  size-dependent fluctuations in the traded value of stocks},
\newblock \bibinfo{journal}{Physical Review E} \bibinfo{volume}{73}
  (\bibinfo{year}{2006}). \DOIprefix\doi{10.1103/physreve.73.046109}.
%Type = Article
\bibitem[{Valenti et~al.(2018)Valenti, Fazio, and Spagnolo}]{Valenti2018}
\bibinfo{author}{D.~Valenti}, \bibinfo{author}{G.~Fazio},
  \bibinfo{author}{B.~Spagnolo},
\newblock \bibinfo{title}{Stabilizing effect of volatility in financial
  markets},
\newblock \bibinfo{journal}{Physical Review E} \bibinfo{volume}{97}
  (\bibinfo{year}{2018}). \DOIprefix\doi{10.1103/physreve.97.062307}.
%Type = Article
\bibitem[{Wang et~al.(2006)Wang, Yamasaki, Havlin, and Stanley}]{Wang2006}
\bibinfo{author}{F.~Wang}, \bibinfo{author}{K.~Yamasaki},
  \bibinfo{author}{S.~Havlin}, \bibinfo{author}{H.~E. Stanley},
\newblock \bibinfo{title}{Scaling and memory of intraday volatility return
  intervals in stock markets},
\newblock \bibinfo{journal}{Physical Review E} \bibinfo{volume}{73}
  (\bibinfo{year}{2006}). \DOIprefix\doi{10.1103/physreve.73.026117}.
%Type = Article
\bibitem[{Hethcote(2000)}]{Hethcote2000}
\bibinfo{author}{H.~W. Hethcote},
\newblock \bibinfo{title}{The mathematics of infectious diseases},
\newblock \bibinfo{journal}{{SIAM} Review} \bibinfo{volume}{42}
  (\bibinfo{year}{2000}) \bibinfo{pages}{599--653}.
  \DOIprefix\doi{10.1137/s0036144500371907}.
%Type = Article
\bibitem[{Chowell et~al.(2016)Chowell, Sattenspiel, Bansal, and
  Viboud}]{Chowell2016}
\bibinfo{author}{G.~Chowell}, \bibinfo{author}{L.~Sattenspiel},
  \bibinfo{author}{S.~Bansal}, \bibinfo{author}{C.~Viboud},
\newblock \bibinfo{title}{Mathematical models to characterize early epidemic
  growth: A review},
\newblock \bibinfo{journal}{Physics of Life Reviews} \bibinfo{volume}{18}
  (\bibinfo{year}{2016}) \bibinfo{pages}{66--97}.
  \DOIprefix\doi{10.1016/j.plrev.2016.07.005}.
%Type = Article
\bibitem[{Machado and Lopes(2020)}]{Machado2020}
\bibinfo{author}{J.~A.~T. Machado}, \bibinfo{author}{A.~M. Lopes},
\newblock \bibinfo{title}{Rare and extreme events: the case of {COVID}-19
  pandemic},
\newblock \bibinfo{journal}{Nonlinear Dynamics}  (\bibinfo{year}{2020}).
  \DOIprefix\doi{10.1007/s11071-020-05680-w}.
%Type = Article
\bibitem[{Basalto et~al.(2007)Basalto, Bellotti, Carlo, Facchi, Pantaleo, and
  Pascazio}]{Basalto2007}
\bibinfo{author}{N.~Basalto}, \bibinfo{author}{R.~Bellotti},
  \bibinfo{author}{F.~D. Carlo}, \bibinfo{author}{P.~Facchi},
  \bibinfo{author}{E.~Pantaleo}, \bibinfo{author}{S.~Pascazio},
\newblock \bibinfo{title}{Hausdorff clustering of financial time series},
\newblock \bibinfo{journal}{Physica A: Statistical Mechanics and its
  Applications} \bibinfo{volume}{379} (\bibinfo{year}{2007})
  \bibinfo{pages}{635--644}. \DOIprefix\doi{10.1016/j.physa.2007.01.011}.
%Type = Article
\bibitem[{Basalto et~al.(2008)Basalto, Bellotti, Carlo, Facchi, Pantaleo, and
  Pascazio}]{Basalto2008}
\bibinfo{author}{N.~Basalto}, \bibinfo{author}{R.~Bellotti},
  \bibinfo{author}{F.~D. Carlo}, \bibinfo{author}{P.~Facchi},
  \bibinfo{author}{E.~Pantaleo}, \bibinfo{author}{S.~Pascazio},
\newblock \bibinfo{title}{Hausdorff clustering},
\newblock \bibinfo{journal}{Physical Review E} \bibinfo{volume}{78}
  (\bibinfo{year}{2008}). \DOIprefix\doi{10.1103/physreve.78.046112}.
%Type = Article
\bibitem[{Mantegna(1999)}]{Mantegna1999}
\bibinfo{author}{R.~Mantegna},
\newblock \bibinfo{title}{Hierarchical structure in financial markets},
\newblock \bibinfo{journal}{The European Physical Journal B}
  \bibinfo{volume}{11} (\bibinfo{year}{1999}) \bibinfo{pages}{193--197}.
  \DOIprefix\doi{10.1007/s100510050929}.
%Type = Article
\bibitem[{Madore et~al.(2007)}]{Madore2007}
\bibinfo{author}{A.-M. Madore}, et~al.,
\newblock \bibinfo{title}{Contribution of hierarchical clustering techniques to
  the modeling of the geographic distribution of genetic polymorphisms
  associated with chronic inflammatory diseases in the {Q}u{\'{e}}bec
  population},
\newblock \bibinfo{journal}{Public Health Genomics} \bibinfo{volume}{10}
  (\bibinfo{year}{2007}) \bibinfo{pages}{218--226}.
  \DOIprefix\doi{10.1159/000106560}.
%Type = Article
\bibitem[{Kretzschmar and Mikolajczyk(2009)}]{Kretzschmar2009}
\bibinfo{author}{M.~Kretzschmar}, \bibinfo{author}{R.~T. Mikolajczyk},
\newblock \bibinfo{title}{Contact profiles in eight {E}uropean countries and
  implications for modelling the spread of airborne infectious diseases},
\newblock \bibinfo{journal}{{PLoS} {ONE}} \bibinfo{volume}{4}
  (\bibinfo{year}{2009}) \bibinfo{pages}{e5931}.
  \DOIprefix\doi{10.1371/journal.pone.0005931}.
%Type = Article
\bibitem[{Alashwal et~al.(2019)Alashwal, Halaby, Crouse, Abdalla, and
  Moustafa}]{Alashwal2019}
\bibinfo{author}{H.~Alashwal}, \bibinfo{author}{M.~E. Halaby},
  \bibinfo{author}{J.~J. Crouse}, \bibinfo{author}{A.~Abdalla},
  \bibinfo{author}{A.~A. Moustafa},
\newblock \bibinfo{title}{The application of unsupervised clustering methods to
  {A}lzheimer's disease},
\newblock \bibinfo{journal}{Frontiers in Computational Neuroscience}
  \bibinfo{volume}{13} (\bibinfo{year}{2019}).
  \DOIprefix\doi{10.3389/fncom.2019.00031}.
%Type = Inproceedings
\bibitem[{Muradi et~al.(2015)Muradi, Bustamam, and Lestari}]{Muradi2015}
\bibinfo{author}{H.~Muradi}, \bibinfo{author}{A.~Bustamam},
  \bibinfo{author}{D.~Lestari},
\newblock \bibinfo{title}{Application of hierarchical clustering ordered
  partitioning and collapsing hybrid in {E}bola virus phylogenetic analysis},
\newblock in: \bibinfo{booktitle}{2015 International Conference on Advanced
  Computer Science and Information Systems ({ICACSIS})},
  \bibinfo{publisher}{{IEEE}}, \bibinfo{year}{2015}, pp.
  \bibinfo{pages}{317--323}. \DOIprefix\doi{10.1109/icacsis.2015.7415183}.
%Type = Article
\bibitem[{Rizzi et~al.(2010)Rizzi, Mahata, Mathieson, and Moscato}]{Rizzi2010}
\bibinfo{author}{R.~Rizzi}, \bibinfo{author}{P.~Mahata},
  \bibinfo{author}{L.~Mathieson}, \bibinfo{author}{P.~Moscato},
\newblock \bibinfo{title}{Hierarchical clustering using the arithmetic-harmonic
  cut: Complexity and experiments},
\newblock \bibinfo{journal}{{PLoS} {ONE}} \bibinfo{volume}{5}
  (\bibinfo{year}{2010}) \bibinfo{pages}{e14067}.
  \DOIprefix\doi{10.1371/journal.pone.0014067}.
%Type = Article
\bibitem[{Ward(1963)}]{Ward1963}
\bibinfo{author}{J.~H. Ward},
\newblock \bibinfo{title}{Hierarchical grouping to optimize an objective
  function},
\newblock \bibinfo{journal}{Journal of the {A}merican Statistical Association}
  \bibinfo{volume}{58} (\bibinfo{year}{1963}) \bibinfo{pages}{236--244}.
  \DOIprefix\doi{10.1080/01621459.1963.10500845}.
%Type = Article
\bibitem[{Szekely and Rizzo(2005)}]{Szekely2005}
\bibinfo{author}{G.~J. Szekely}, \bibinfo{author}{M.~L. Rizzo},
\newblock \bibinfo{title}{Hierarchical clustering via joint between-within
  distances: Extending {W}ard's minimum variance method},
\newblock \bibinfo{journal}{Journal of Classification} \bibinfo{volume}{22}
  (\bibinfo{year}{2005}) \bibinfo{pages}{151--183}.
  \DOIprefix\doi{10.1007/s00357-005-0012-9}.
%Type = Article
\bibitem[{Wang and Song(2011)}]{Wang2011}
\bibinfo{author}{H.~Wang}, \bibinfo{author}{M.~Song},
\newblock \bibinfo{title}{Ckmeans.1d.dp: Optimal k-means clustering in one
  dimension by dynamic programming},
\newblock \bibinfo{journal}{The {R} Journal} \bibinfo{volume}{3}
  (\bibinfo{year}{2011}) \bibinfo{pages}{29--33}.
  \DOIprefix\doi{10.32614/rj-2011-015}.
%Type = Article
\bibitem[{Radchenko and Mukherjee(2017)}]{Radchenko2017}
\bibinfo{author}{P.~Radchenko}, \bibinfo{author}{G.~Mukherjee},
\newblock \bibinfo{title}{Convex clustering via $\ell_1$ fusion penalization},
\newblock \bibinfo{journal}{Journal of the Royal Statistical Society: Series B
  (Statistical Methodology)} \bibinfo{volume}{79} (\bibinfo{year}{2017})
  \bibinfo{pages}{1527--1546}. \DOIprefix\doi{10.1111/rssb.12226}.
%Type = Article
\bibitem[{Aminikhanghahi and Cook(2016)}]{Aminikhanghahi2016}
\bibinfo{author}{S.~Aminikhanghahi}, \bibinfo{author}{D.~J. Cook},
\newblock \bibinfo{title}{A survey of methods for time series change point
  detection},
\newblock \bibinfo{journal}{Knowledge and Information Systems}
  \bibinfo{volume}{51} (\bibinfo{year}{2016}) \bibinfo{pages}{339--367}.
  \DOIprefix\doi{10.1007/s10115-016-0987-z}.
%Type = Article
\bibitem[{Lavielle and Teyssi{\`{e}}re(2006)}]{Lavielle2006}
\bibinfo{author}{M.~Lavielle}, \bibinfo{author}{G.~Teyssi{\`{e}}re},
\newblock \bibinfo{title}{Detection of multiple change-points in multivariate
  time series},
\newblock \bibinfo{journal}{Lithuanian Mathematical Journal}
  \bibinfo{volume}{46} (\bibinfo{year}{2006}) \bibinfo{pages}{287--306}.
  \DOIprefix\doi{10.1007/s10986-006-0028-9}.
%Type = Article
\bibitem[{James and Menzies(2020)}]{james2020covidusa}
\bibinfo{author}{N.~James}, \bibinfo{author}{M.~Menzies},
\newblock \bibinfo{title}{{COVID}-19 in the {United States}: Trajectories and
  second surge behavior},
\newblock \bibinfo{journal}{Chaos: An Interdisciplinary Journal of Nonlinear
  Science} \bibinfo{volume}{30} (\bibinfo{year}{2020}) \bibinfo{pages}{091102}.
  \DOIprefix\doi{10.1063/5.0024204}.
%Type = Article
\bibitem[{James et~al.(2020)James, Menzies, Azizi, and Chan}]{James2020_nsm}
\bibinfo{author}{N.~James}, \bibinfo{author}{M.~Menzies},
  \bibinfo{author}{L.~Azizi}, \bibinfo{author}{J.~Chan},
\newblock \bibinfo{title}{Novel semi-metrics for multivariate change point
  analysis and anomaly detection},
\newblock \bibinfo{journal}{Physica D: Nonlinear Phenomena}
  \bibinfo{volume}{412} (\bibinfo{year}{2020}) \bibinfo{pages}{132636}.
  \DOIprefix\doi{10.1016/j.physd.2020.132636}.
%Type = Misc
\bibitem[{wor(2020)}]{worldbankgdp}
\bibinfo{title}{{GDP} (current {US}\$)},
  \bibinfo{howpublished}{\url{https://data.worldbank.org/indicator/NY.GDP.MKTP.CD?year_high_desc=true}},
  \bibinfo{year}{2020}. \bibinfo{note}{{The World Bank}, September 21, 2020}.
%Type = Misc
\bibitem[{Arouxet et~al.(2020)Arouxet, Bariviera, Pastor, and
  Vampa}]{arouxet2020}
\bibinfo{author}{M.~B. Arouxet}, \bibinfo{author}{A.~F. Bariviera},
  \bibinfo{author}{V.~E. Pastor}, \bibinfo{author}{V.~Vampa},
  \bibinfo{title}{Covid-19 impact on cryptocurrencies: evidence from a
  wavelet-based {H}urst exponent}, \bibinfo{year}{2020}.
  \href{http://arxiv.org/abs/2009.05652}{{\tt arXiv:2009.05652}}.
%Type = Book
\bibitem[{Rudin(1991)}]{RudinFA}
\bibinfo{author}{W.~Rudin}, \bibinfo{title}{Functional {A}nalysis},
  \bibinfo{publisher}{McGraw-Hill Science}, \bibinfo{year}{1991}.
%Type = Misc
\bibitem[{Boadle(2020)}]{Reutersbrazil}
\bibinfo{author}{A.~Boadle}, \bibinfo{title}{Brazil has record new coronavirus
  cases, surpasses {F}rance in deaths},
  \bibinfo{howpublished}{\url{https://www.reuters.com/article/us-health-coronavirus-brazil/brazils-coronavirus-outbreak-worsens-as-total-cases-near-500000-idUSKBN2360U8}},
  \bibinfo{year}{2020}. \bibinfo{note}{{Reuters}, May 31, 2020}.
%Type = Misc
\bibitem[{Neuman(2020{\natexlab{a}})}]{NPRJune}
\bibinfo{author}{S.~Neuman}, \bibinfo{title}{France announces further reopening
  amid declining number of coronavirus cases},
  \bibinfo{howpublished}{\url{https://www.npr.org/sections/coronavirus-live-updates/2020/06/15/876953360/france-announces-further-reopening-amid-declining-number-of-coronavirus-cases}},
  \bibinfo{year}{2020}{\natexlab{a}}. \bibinfo{note}{{NPR}, June 15, 2020}.
%Type = Misc
\bibitem[{Neuman(2020{\natexlab{b}})}]{SMHAug}
\bibinfo{author}{S.~Neuman}, \bibinfo{title}{Global coronavirus cases hit 20
  million as pandemic accelerates},
  \bibinfo{howpublished}{\url{https://www.smh.com.au/world/europe/who-only-10-per-cent-of-the-way-to-funding-coronavirus-fight-20200811-p55kfz.html}},
  \bibinfo{year}{2020}{\natexlab{b}}. \bibinfo{note}{{Sydney Morning Herald},
  August 11, 2020}.
%Type = Misc
\bibitem[{McDonell(2020)}]{australiabbc_2020}
\bibinfo{author}{S.~McDonell}, \bibinfo{title}{Coronavirus: {US} and
  {A}ustralia close borders to {C}hinese arrivals},
  \bibinfo{howpublished}{\url{https://www.bbc.com/news/world-51338899}},
  \bibinfo{year}{2020}. \bibinfo{note}{{BBC}, Accessed February 1, 2020}.
%Type = Misc
\bibitem[{McCurry(2020)}]{Koreaguardian_2020}
\bibinfo{author}{J.~McCurry}, \bibinfo{title}{Test, trace, contain: how {S}outh
  {K}orea flattened its coronavirus curve},
  \bibinfo{howpublished}{\url{https://www.theguardian.com/world/2020/apr/23/test-trace-contain-how-south-korea-flattened-its-coronavirus-curve}},
  \bibinfo{year}{2020}. \bibinfo{note}{{The Guardian}, April 23, 2020}.
%Type = Misc
\bibitem[{McCann et~al.(2020)McCann, Popovich, and Wu}]{italynyt2020}
\bibinfo{author}{A.~McCann}, \bibinfo{author}{N.~Popovich},
  \bibinfo{author}{J.~Wu}, \bibinfo{title}{Italy’s virus shutdown came too
  late. what happens now?},
  \bibinfo{howpublished}{\url{https://www.nytimes.com/interactive/2020/04/05/world/europe/italy-coronavirus-lockdown-reopen.html}},
  \bibinfo{year}{2020}. \bibinfo{note}{{The New York Times}, April 5, 2020}.
%Type = Article
\bibitem[{Goodell and Goutte(2020)}]{Goodell2020}
\bibinfo{author}{J.~W. Goodell}, \bibinfo{author}{S.~Goutte},
\newblock \bibinfo{title}{Co-movement of {COVID}-19 and {B}itcoin: Evidence
  from wavelet coherence analysis},
\newblock \bibinfo{journal}{Finance Research Letters}  (\bibinfo{year}{2020})
  \bibinfo{pages}{101625}. \DOIprefix\doi{10.1016/j.frl.2020.101625}.
%Type = Article
\bibitem[{James et~al.(2021)James, Menzies, and Chan}]{James2021_crypto}
\bibinfo{author}{N.~James}, \bibinfo{author}{M.~Menzies},
  \bibinfo{author}{J.~Chan},
\newblock \bibinfo{title}{Changes to the extreme and erratic behaviour of
  cryptocurrencies during {COVID}-19},
\newblock \bibinfo{journal}{Physica A: Statistical Mechanics and its
  Applications} \bibinfo{volume}{565} (\bibinfo{year}{2021})
  \bibinfo{pages}{125581}. \DOIprefix\doi{10.1016/j.physa.2020.125581}.
%Type = Misc
\bibitem[{wor(2020)}]{worldindata2020}
\bibinfo{title}{Our {W}orld in {D}ata},
  \bibinfo{howpublished}{\url{https://ourworldindata.org/coronavirus-source-data}},
  \bibinfo{year}{2020}. \bibinfo{note}{Accessed September 6, 2020}.
%Type = Article
\bibitem[{Milligan(1980)}]{Milligan1980}
\bibinfo{author}{G.~W. Milligan},
\newblock \bibinfo{title}{An examination of the effect of six types of error
  perturbation on fifteen clustering algorithms},
\newblock \bibinfo{journal}{Psychometrika} \bibinfo{volume}{45}
  (\bibinfo{year}{1980}) \bibinfo{pages}{325--342}.
  \DOIprefix\doi{10.1007/bf02293907}.
%Type = Article
\bibitem[{Rousseeuw(1987)}]{Rousseeuw1987}
\bibinfo{author}{P.~J. Rousseeuw},
\newblock \bibinfo{title}{Silhouettes: A graphical aid to the interpretation
  and validation of cluster analysis},
\newblock \bibinfo{journal}{Journal of Computational and Applied Mathematics}
  \bibinfo{volume}{20} (\bibinfo{year}{1987}) \bibinfo{pages}{53--65}.
  \DOIprefix\doi{10.1016/0377-0427(87)90125-7}.
%Type = Article
\bibitem[{Krzanowski and Lai(1988)}]{krzanowski1988}
\bibinfo{author}{W.~J. Krzanowski}, \bibinfo{author}{Y.~T. Lai},
\newblock \bibinfo{title}{A criterion for determining the number of groups in a
  data set using sum-of-squares clustering},
\newblock \bibinfo{journal}{Biometrics} \bibinfo{volume}{44}
  (\bibinfo{year}{1988}) \bibinfo{pages}{23--34}.
  \DOIprefix\doi{10.2307/2531893}.
%Type = Article
\bibitem[{Hubert and Levin(1976)}]{Hubert1976}
\bibinfo{author}{L.~J. Hubert}, \bibinfo{author}{J.~R. Levin},
\newblock \bibinfo{title}{A general statistical framework for assessing
  categorical clustering in free recall.},
\newblock \bibinfo{journal}{Psychological Bulletin} \bibinfo{volume}{83}
  (\bibinfo{year}{1976}) \bibinfo{pages}{1072--1080}.
  \DOIprefix\doi{10.1037/0033-2909.83.6.1072}.
%Type = Article
\bibitem[{McClain and Rao(1975)}]{Mcclain1975}
\bibinfo{author}{J.~O. McClain}, \bibinfo{author}{V.~R. Rao},
\newblock \bibinfo{title}{{CLUSTISZ}: A program to test for the quality of
  clustering of a set of objects},
\newblock \bibinfo{journal}{Journal of Marketing Research} \bibinfo{volume}{12}
  (\bibinfo{year}{1975}) \bibinfo{pages}{456--460}.
%Type = Article
\bibitem[{Dunn(1974)}]{Dunn1974}
\bibinfo{author}{J.~C. Dunn},
\newblock \bibinfo{title}{Well-separated clusters and optimal fuzzy
  partitions},
\newblock \bibinfo{journal}{Journal of Cybernetics} \bibinfo{volume}{4}
  (\bibinfo{year}{1974}) \bibinfo{pages}{95--104}.
  \DOIprefix\doi{10.1080/01969727408546059}.
%Type = Article
\bibitem[{Lauer et~al.(2020)}]{incubation2020}
\bibinfo{author}{S.~A. Lauer}, et~al.,
\newblock \bibinfo{title}{The incubation period of {C}oronavirus disease 2019
  ({COVID}-19) from publicly reported confirmed cases: Estimation and
  application},
\newblock \bibinfo{journal}{Annals of Internal Medicine} \bibinfo{volume}{172}
  (\bibinfo{year}{2020}) \bibinfo{pages}{577--582}.
  \DOIprefix\doi{10.7326/m20-0504}.

\end{thebibliography}
%\end{nolinenumbers}
\end{document}